\documentclass{aa}  
\usepackage{multirow} 
\usepackage{placeins}
\usepackage{graphicx}
\usepackage{txfonts}
\usepackage{lipsum}
\usepackage{subcaption}                                         
\usepackage{lscape}                                             
\usepackage{placeins}                                       
\usepackage{hyperref}

\hypersetup{
    colorlinks=true,
    linkcolor=blue,
    filecolor=blue,      
    urlcolor=blue,
    citecolor=blue
    }

\newcommand{\linkorcid}[1]{\href{https://orcid.org/#1}{\includegraphics[width=8pt]{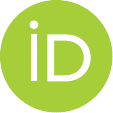}}}

\begin{document}

   \title{ Characterising a population of fast-declining SNe Ia in galaxy cluster cores}
    \authorrunning{A.~Eranhalodi et al.}
    \titlerunning{ SNe Ia in galaxy clusters cores}

   \author
   {A.~Eranhalodi\inst{\ref{hu}, \ref{desy}}\fnmsep\thanks{Corresponding author; akshay.eranhalodi@desy.de}~\linkorcid{0000-0002-2958-0219} ,
    J.~Nordin\inst{\ref{hu}} \linkorcid{0000-0001-8342-6274} \and
    M.~Kowalski\inst{\ref{hu},\ref{desy}}~\linkorcid{0000-0001-8594-8666} \and
    U.~Burgaz \inst{\ref{tcd}}~\linkorcid{0000-0003-0126-3999 } \and 
    F.~Fischer \inst{\ref{hu}, \ref{desy}}~\linkorcid{0009-0002-7161-7473} \and 
    M.~Ginolin \inst{\ref{cmb}}~\linkorcid{0009-0004-5311-9301} \and 
    Y.-L.~Kim \inst{\ref{ylm}}~\linkorcid{0000-0002-1031-0796} \and 
    T.~E.~M\"uller-Bravo\inst{\ref{tcd}, \ref{bravo2}}~\linkorcid{0000-0003-3939-7167} \and
    J.~Sollerman\inst{\ref{okc}}~\linkorcid{0000-0003-1546-6615} \and
    A.~Townsend \inst{\ref{alice}}~\linkorcid{0000-0001-6343-3362} \and
    A.~Drake \inst{\ref{cahill}} \and 
    M.~Graham \inst{\ref{graham}}~\linkorcid{0000-0002-3168-0139} \and
    D.~Hale \inst{\ref{coob}} \and
    J.~C.~Jaimes \inst{\ref{graham}}~\linkorcid{0000-0002-0987-3372} \and
    R.~R.~Laher \inst{\ref{ipac}}~\linkorcid{0000-0003-2451-5482} \and
    J.~Purdum \inst{\ref{cahill}}~\linkorcid{0000-0003-1227-3738} 
    }

   \institute
   {Institut f\"ur Physik, Humboldt-Universit\"at zu Berlin, Newtonstr. 15, 12489 Berlin, Germany \label{hu} 
    \and
    Deutsches Elektronen-Synchrotron, 15735 Zeuthen, Germany \label{desy} \and
    School of Physics, Trinity College Dublin, College Green, Dublin 2, Ireland \label{tcd} 
    \and
    Institute of Astronomy and Kavli Institute for Cosmology, University of Cambridge, Madingley Road, Cambridge CB3 0HA, UK \label{cmb}
    \and
    Department of Astronomy \& Center for Galaxy Evolution Research, Yonsei University, Seoul 03722, Republic of Korea \label{ylm}
    \and
    Instituto de Ciencias Exactas y Naturales (ICEN), Universidad Arturo Prat, Chile \label{bravo2}
    \and
    School of Physics and Astronomy, University of Birmingham, Birmingham B15 2TT, UK \label{alice}
    \and
    Oskar Klein Centre, Department of Physics, Stockholm University, Albanova University Center, SE 106 91 Stockholm, Sweden \label{okc}
    \and
    Caltech Optical Observatories, California Institute of Technology, Pasadena, CA  91125 \label{coob}
    \and
     Cahill Center for Astrophysics, California Institute of Technology, 1200 E. California Boulevard, Pasadena, CA 91125, USA \label{cahill}
    \and
    Division of Physics, Mathematics and Astronomy, California Institute of Technology, Pasadena, CA 91125, USA \label{graham}
    \and 
    IPAC, California Institute of Technology, 1200 E. California Blvd, Pasadena, CA 91125, USA \label{ipac}
    }

   \date{Received October 28, 2026}

  \abstract
   { Galaxy cluster cores, characterised by a dominant old stellar population and suppressed star-formation rates, are ideal for studying the environmental dependence of Type Ia supernovae (SNe Ia). It is unclear whether the diversity of SNe Ia, parametrised by light curve stretch, stems from distinct explosion mechanisms or varying initial conditions, both closely tied to progenitor age.}
   {  We analyse the light curve  stretch and colour distribution of photometrically identified SNe Ia, from the oldest  progenitor population within galaxy clusters cores, utilising the Zwicky Transient Facility (ZTF) archival data.} 
   { Using the AMPEL platform, we filter 89,025 ZTF transients spatially coincident with our galaxy cluster catalogues to identify 4,969 likely SN Ia candidates.  We split this sample into three populations: core cluster SNe Ia  ($r \leq 0.5R_{500}$), outer-core cluster SNe Ia ($0.5R_{500}< r\leq 2R_{500}$), and a reference population of non-cluster SN Ia, which serves as a field comparison sample.  }
  {We present the largest sample of photometrically identified cluster SNe Ia to date, notably isolating a population of 242 SNe Ia within innermost cluster cores ($ r \leq 0.5R_{500}$). This fast-declining SN Ia population is statistically different from the outer-core cluster and field SN Ia population at $\sim4\sigma$ level. A marginal statistical preference for a bimodal distribution over a unimodal one ($\Delta AIC$ of 1.53 and 2.3 for each catalogue), indicate cluster cores almost entirely host fast-declining SNe Ia. Conversely, the colour distribution show no significant environmental variation.}
   {By filtering out the prompt SN Ia population, cluster cores provide an environment to directly probe the delayed SN Ia progenitor channel. Our observations support a two-population model where progenitor age drives the light curve stretch diversity. The unique conditions in cluster cores combined with the low-stretch value observed in our sample, suggest sub-Chandrasekhar mass double degenerate systems as progenitors of core cluster SNe Ia.}

   \keywords{supernovae: general - Galaxies: clusters: general     }
   \maketitle
   \nolinenumbers
   
\newcommand{\x}{$x_1$}

\section{Introduction}

A  Type Ia supernova (SN Ia) is a specific class of stellar explosion occurring in a binary system involving at least one carbon-oxygen white dwarf \citep{hoyle1960nucleosynthesis}. The explosion is triggered by the white dwarf's interaction with its companion, either through accretion from a non-degenerate star or a merger with another white dwarf. This interaction initiates a thermonuclear runaway, which completely disrupts the white dwarf and produces the observed SN Ia. The intrinsic diversity of the majority of observed SNe Ia can be corrected by a relationship between their peak luminosity and their photometric light curve  properties \citep{phillips1993absolute, hamuy1996morphology, 1998A&A...331..815T}, specifically the stretch (\x) and colour ($c$) \citep{saltguy}. Applying these corrections reduces the dispersion in SN Ia brightness from $\approx$ 0.40 mag to $\approx$ 0.15 mag, allowing distances to SNe Ia to be estimated within $\approx 6\%$ accuracy. This standardisable nature of SNe Ia enabled the fundamental discovery of the accelerated expansion of the universe \citep{riess1998observational, perlmutter1999measurements} and have rendered SNe Ia as the most important cosmological distance indicator. 

As crucial SNe Ia have been to cosmology, several open questions remain regarding their progenitor systems, evolutionary pathways and explosion mechanisms. Currently accepted models for the progenitor systems fall into two categories: the single degenerate (SD) model \citep{whelan1973binaries} where the companion is a non-degenerate star, typically a main-sequence or red giant, that facilitates mass transfer via Roche lobe overflow, and the double degenerate (DD) model \citep{webbink1984double}, where a binary white dwarf system evolves through the emission of gravitational waves, leading to orbital decay and the eventual merger of the two compact objects. The fundamental assumption behind using SNe Ia as cosmological distance indicators is that they remain standardisable candles across redshift and progenitor environment. A strategy to validate this assumption and investigate any environmental effect of SN Ia properties is to look for SNe Ia in unique environments such as galaxy clusters. Galaxy clusters are predominantly composed of early-type elliptical galaxies characterised by minimal star formation and low dust content \citep{draine1979physics, mcgee2010dust}. Furthermore, the cores of galaxy clusters exhibit distinct physical properties compared to their outer regions and field environments, such as cluster cores ($\leq0.5R_{500}$) being more metal-rich than rest of the clusters \citep{mantz2017metallicity, romero2023sami}, and galaxies near the virial radius ($1\,R_{200}$) featuring younger mean stellar population ages (by $15 \pm 4\%$) relative to core galaxies \citep{Cluster_age1, Cluster_age2}. Consequently, evaluating SNe Ia partitioned across cluster cores, outer cores, and the field provides an ideal test bed to investigate the progenitor systems and host environment dependence of SN Ia properties. 

A fundamental approach to constrain the progenitor scenarios involve measuring their delay time distribution (DTD). The DTD traces the SN Ia rate over time that would follow a brief burst of star formation, with different progenitor scenarios predicting different DTDs. The field SN Ia rate measurements suggest the existence of two SN Ia populations following a burst of star formation, a prompt population that explodes within $\approx 100-500$ Myr, and a delayed channel that produces SNe Ia on timescales in order of 5 Gyr \citep{2005A&A...433..807M, 2006ApJ...648..868S, 2005ApJ...629L..85S, 2012ApJ...755...61S}. In galaxy clusters, the stellar population is assumed to have been formed in a single brief burst at a redshift of $z \approx 3$ \citep{Eisenhardt2008clusters, Kurk2009clusters, Mancone2009clusters}. This simple star-formation history (SFH) allows for a more accurate derivation of the DTD, compared to the uncertainties inherent in determining the complex SFH of field galaxies. Assuming an instantaneous star-formation burst at $z_f = 3$, the SNe Ia rate in galaxy clusters peaks at short delay times ($\leq5 \text{Gyr}$, corresponding to $1.45 \leq z \leq 3$), and at longer delay times ($\geq5 \text{Gyr}$, corresponding to $z\leq1.45$) the rate almost remains a constant (see Fig. 2 in \citealt{maoz2010supernova}). Furthermore, the cluster DTD analysis by \cite{maoz2010supernova} demonstrates that pure SD or DD model yields poor fit to the observed DTD. Instead, the observed rate can be reproduced by a combination of the prompt SD and the delayed channel, under which assumption the vast majority of SNe Ia at $z<1.5$  originate from DD systems. If this scenario holds, studying SNe Ia in galaxy clusters in the nearby redshift regime could potentially isolate the delayed progenitor channel. 

The properties of SNe Ia are known to depend on their environments \citep[e.g.,][]{hamuy1996morphology, hamuy2000search, Sullivan2006,  neill2009local, Lampeitl}. Even after the initial  SN Ia luminosity standardisation with light curve stretch and colour, the residual scatter shows correlation with environmental properties \citep[e.g. ][]{Kelly2010ApJ, sullivan2010dependence, childress2013host, betoule2014improved, Scolnic2018, smith2020first}. This correlation is now used as a mass-step correction in addition to the light curve stretch and colour corrections  in cosmological analyses, with the step typically defined around host galaxy mass of $\text{M}_{\star}\sim10^{10} \, \text{M}_{\odot}$.  The physical origin of this mass-step remains highly  ambiguous, and since the mass of a galaxy correlates with other galaxy properties such as its metallicity, star-formation rate and age, there have been several attempts in the literature to explain the mass step driven by these local environmental indicators (within $1-3$ kpc of the SN Ia site) \citep[e.g.,][]{rigault2013evidence, rigault2020strong, kim2018environmental, kelsey2023concerning, meldorf2023dark, Chung_age_bias_25}. As these global and local host galaxy properties would evolve with redshift we should speculate if SNe Ia standardisation relations are also redshift dependent. For example, \cite{nicolas2021redshift} identified evidence of redshift evolution of the stretch parameter and introduced a redshift drift model based on predictions from \cite{rigault2020strong}. They find the stretch distribution is inherently bimodal, featuring a low-stretch mode exclusively associated with old environments, and a high-stretch mode that is shared between young and old environments. Furthermore, growing evidences suggest that the corrections utilised for SN Ia standardisation may be more complex than the currently implemented linear trends, potentially varying across environments and progenitor channels \citep{ginolin2025ztf, wojtak2025stretch, rubin2026banana}. So in conclusion understanding environmental dependencies of SN Ia properties is crucial to maintaining their efficacy as standardisable candles. This is particularly interesting while studying SN Ia in galaxy clusters as it can isolate a population that could be used to study and untangle these effects.

Prior works on SNe Ia in galaxy clusters have found that their properties systematically differ from those in field environments, with cluster SNe Ia showing faster decline rates (low-stretch) compared to field SNe Ia \citep{meyers2012hubble, xavier2013properties, toy2023rates, larison2024environmental, ruppin2025ztf}. In contrast, these studies find no significant differences in colour between cluster and field SNe Ia. \cite{ruppin2025ztf}, building on the redshift-drift model introduced by \cite{nicolas2021redshift}, added a new spatial parameter on how the SN Ia stretch evolves continuously as a function of projected distance to the nearest galaxy cluster, using the data from second data release (DR2) of the Zwicky Transient Facility (ZTF) \citep{2025A&A...694A...1R}. \cite{larison2024environmental} showed that when dividing their cluster SNe Ia sample into an inner ($r_{500}$) and outer sample ($r_{500}-2r_{500}$), the inner cluster sample contains a significantly higher fraction of fast evolving SNe Ia than both the outer cluster sample and field galaxies. Conversely, \cite{toy2023rates} observed no significant difference in the SN Ia light curve  properties between field and cluster SNe Ia when the comparison is limited to host galaxies with similar mass and colour. While the findings of \cite{toy2023rates}  suggest that the local host galaxy properties rather than the unique cluster environment drive the cluster-field SN Ia light curve  variations, this discrepancy may stem from the higher redshift range of their cluster SNe Ia sample ($0.1\leq z\leq0.7$), compared to the low redshift range probed by \cite{ruppin2025ztf} ($z\leq0.07$) and \cite{larison2024environmental} ($z\leq0.1$).

Despite observed differences between cluster and field SNe Ia, it remains unclear whether these trends arise from distinct explosion mechanisms or different progenitor channels. Furthermore, existing constraints from cluster SNe Ia are limited by relatively small sample sizes. For instance, \cite{xavier2013properties} and \cite{toy2023rates} identified 48 and 66 SNe Ia within 1.5 Mpc of cluster centres, while \cite{larison2024environmental} utilised a sample of 102 cluster SNe Ia (54 within $r_{500}$ and 48 within $r_{500}-2r_{500}$). Most recently, \cite{ruppin2025ztf} identified 129 SNe Ia within $5r_{500}$ radius. In this work, we revisit the environmental dependence of SN Ia stretch and colour by conducting a large scale population study focused on galaxy cluster cores ($R<0.5R_{500}$). By targeting these innermost regions, we can create a representative sample of SNe Ia originating from  the oldest progenitor population. While prior studies have determined that the global SN Ia stretch distribution is best described by two populations, the scarcity of observed events have made it difficult to isolate them. By leveraging the wide field ZTF survey and employing photometric identification of SNe Ia, we can achieve a significantly larger and deeper sample providing the statistical power necessary to constrain the environmental dependence of SN Ia properties. We analyse SN Ia properties as a function of cluster-centric distance (distance to the cluster scaled by the cluster radius). While potential contamination by non-SN Ia transients might dilute the statistical significance of our findings, this noise is expected to be independent of the distance from the cluster centre and would not introduce a spurious signal.

This paper is organised as follows: Sect.  \ref{sec:ztf} introduces the wide-field ZTF survey used to curate the sample of SNe Ia. Section  \ref{sec:cat} presents the two galaxy cluster catalogues used in this analysis. The analytical pipeline, methodology, and cluster-SN Ia matching procedure are detailed in Sect. \ref{sec:methods}. Finally, Sect.  \ref{sec:prop} presents the observed properties of cluster SNe Ia, followed by a broader discussion and conclusion of our findings in Sect. \ref{sec:disc} and \ref{sec:conclude}.

\section{Transient discovery with wide field surveys} \label{sec:ztf}

\begin{figure*}[!ht]
    \centering
    \includegraphics[width=0.29\linewidth]{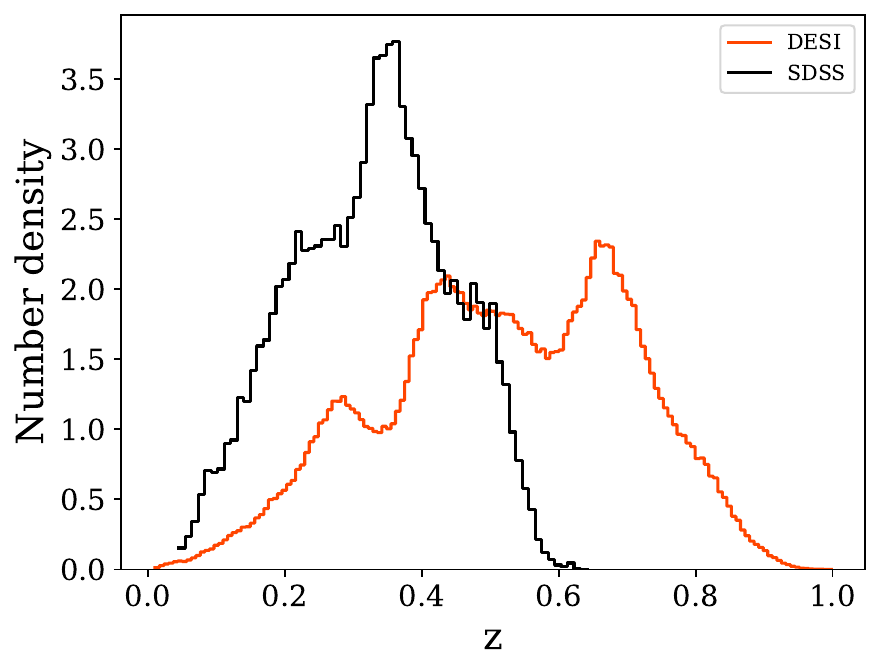}
    \includegraphics[width=0.29\linewidth]{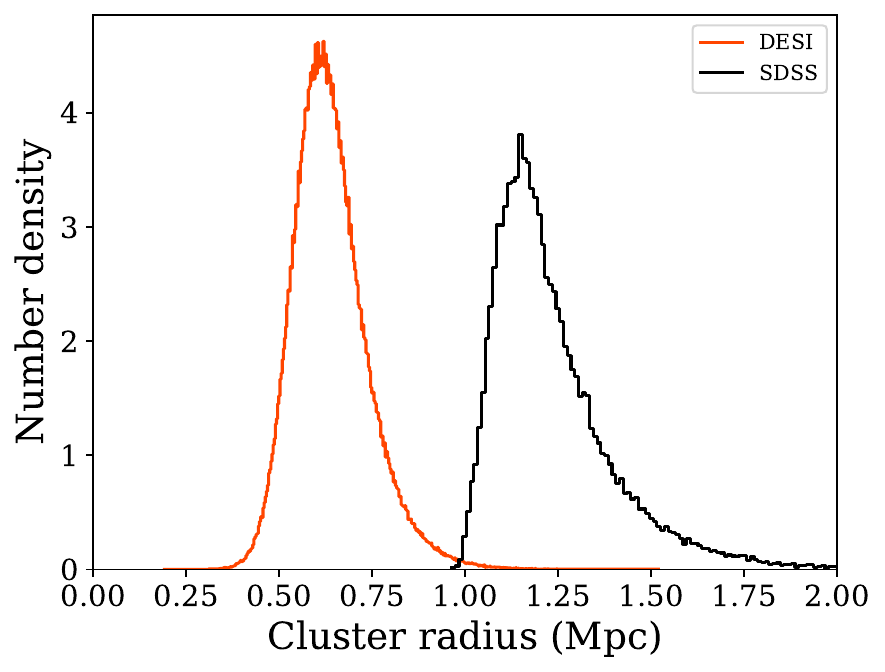}
    \includegraphics[width=0.29\linewidth]{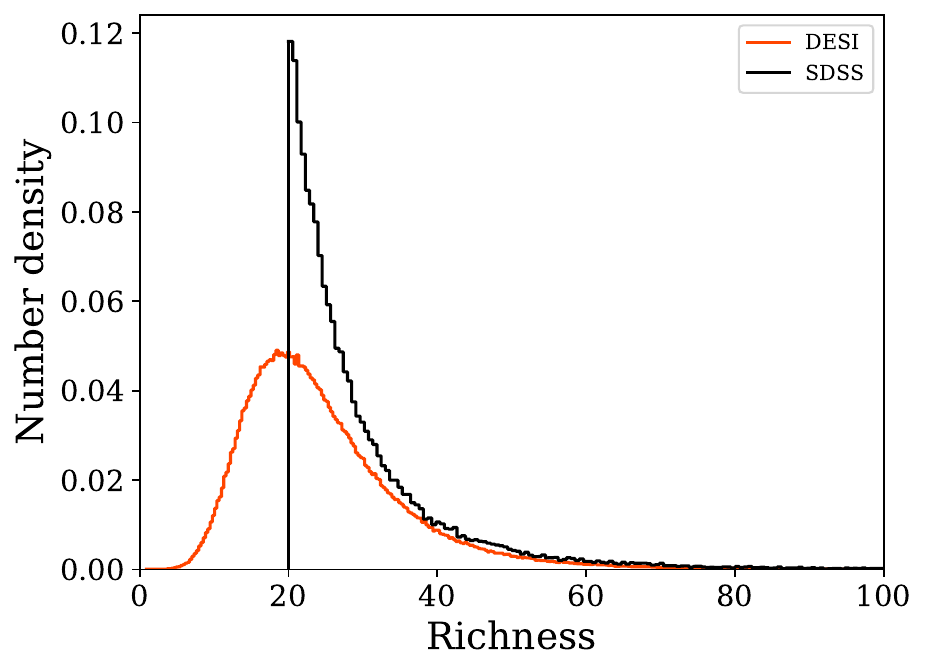}
    \caption{Redshift distribution, cluster radius ($R_{200}$ for SDSS cluster catalogue, $R_{500}$ for DESI cluster catalogue) and richness distribution of galaxy clusters from the two catalogues used in this study. }
    \label{fig:cluster-prop}
\end{figure*}

\subsection{Zwicky Transient Facility}
ZTF is an optical time-domain survey designed to systematically explore the transient and variable sky, operational since 2018. It builds upon the legacy of its predecessor surveys, the Palomar Transient Factory \citep[PTF,][]{law2009palomar} and the intermediate Palomar Transient Factory \citep[iPTF,][]{Kulkarni2013}. The survey operates on the 48-inch Samuel Oschin Schmidt Telescope at the Palomar Observatory equipped with a 600 megapixel mosaic CCD camera that provides a massive field of view of 47 deg$^2$ \citep{2020PASP..132c8001D}. This allows ZTF to scan the entire northern visible sky at an incredibly fast rate of approximately 3,750 deg$^2$ per hour, reaching median depths of around 20.8 magnitude in the $g$ band and 20.6 magnitude in the $r$ band during a nominal 30 second exposure \citep{bellm2019zwicky, graham2019zwicky, masci2018zwicky}. 

ZTF's data processing pipeline can identify transient events and generate alert packets typically within 13 minutes of observation \citep{masci2018zwicky}, streaming up to 1.2 million alert packets nightly via Kafka \citep{Patterson_ztf}. This rapid detection is complemented by the ZTF Bright Transient Survey (BTS) program utilising a dedicated low-resolution integral field spectrograph \citep[SEDm,][]{Blagorodnova, Rigault2019sedm} to spectroscopically classify transients brighter than $18.5$ magnitude in \emph{g} or \emph{r} ZTF filters \citep{2020ApJ...895...32F, 2020ApJ...904...35P}. This combination of wide-field, high-cadence imaging and rapid spectroscopic follow up has allowed ZTF to drastically increase the number of known and well characterised transients. This is illustrated by the ZTF DR2 which contains 3,628 spectroscopically confirmed SNe Ia, increasing the number of low-redshift ($z < 0.3$) SNe Ia by an order of magnitude \citep{2025A&A...694A...1R}. This number is expected to grow even further with the full ZTF survey covering 2018 to 2025 to yield nearly 10,000 spectroscopically confirmed SNe Ia. 

Despite these staggering numbers, the ZTF archive still contains a wealth of unclassified transients. Estimates suggest close to 40,000 additional SNe Ia in the archive identifiable through their photometric properties, that could be used for future analyses. Exploiting this archival potential to photometrically identify likely SN Ia candidates within galaxy clusters is one of the core objectives of this work. 

\subsection{Data irreversibility}
Modern wide field astronomical surveys, such as ZTF, generate vast volumes of data, with the transient detection rates projected to rise by another order of magnitude with next generation surveys such as the Legacy Survey of Space and Time (LSST) at the Vera Rubin Observatory \citep{ivezic2019lsst}. For these transient discovery engines, objects of interest are selected from the real time stream and immediate reactions are requested based on pre-existing criteria (filters). This reaction is usually a follow-up observation using a spectrograph to classify the transient. However, due to limited observational resources and brief reaction windows, only a small fraction are spectroscopically confirmed. In addition, considering that there are likely multiple, parallel and evolving selection criteria for what to follow up on, leads to an irreversible information loss in the archive. As a result, a vast number of alerts remain permanently unclassified. In the context of our analysis, this translates to a significant loss of potential cluster SNe Ia that lack both spectroscopic classification and host galaxy redshift. The ZTF survey have been operational since 2018 and the archive would have  accumulated many unclassified cluster SNe Ia. By identifying these candidates based purely on their optical photometric light curves, we can bypass the bottlenecks of limited spectroscopic classifications. This allows us to transition from a small sample of well-characterised objects to statistics offered by a much larger dataset. By exploring different selection criteria on the ZTF archival data, we tune a filter that maximises the scientific potential of the archive by maximising the confirmed number of SNe Ia from galaxy clusters. This approach of using a static archive to emulate and optimise real-time filters for both computational efficiency and scientific potential exemplifies the dynamic archive concept \citep{redelbach_2024_10692169} established by the \cite{PUNCH4NFDI2020}\footnote{\href{https://www.punch4nfdi.de}{https://www.punch4nfdi.de}}.

\section{Galaxy cluster catalogues} \label{sec:cat}

For this analysis we use two existing galaxy cluster catalogues, one compiled from the data release 9 of the Sloan Digital Sky Survey (SDSS) \citep{banerjee2018optical} and another based on the Dark Energy Spectroscopic Instrument (DESI) legacy imaging surveys \citep{zou2021galaxy}. These catalogues are selected based on two major criteria, 
\begin{itemize}
    \item Maximise the catalogue overlap with the ZTF survey footprint (entire sky north of declination $\sim -31^{\circ}$, \citealt{2019JPhCS1231a2010N} ). 
    \item Maximise the number of identified galaxy clusters available for SN Ia association.
\end{itemize}
Figure \ref{fig:projection} shows the Mollweide projection of the galaxy clusters from the catalogues we used, which is in excellent overlap with the ZTF sky. Galaxy clusters can be identified through their various signatures across the electromagnetic spectrum. Galaxy cluster catalogues identified through their X-ray emission and the Sunyaev–Zel’dovich effect generally yield high-purity samples, but are biased towards high-mass systems. In contrast, cluster catalogues that rely on optical and near-infrared signatures produce much larger samples, but these can suffer from lower purity due to chance associations with field galaxies. Since this study aims to maximise the number of identified galaxy cluster SNe Ia both the catalogues selected are constructed using optical and infrared observations.

\subsection{SDSS galaxy cluster catalogue}
The SDSS galaxy cluster catalogue (SGCC) is an optical catalogue of galaxy clusters compiled from the data release 9 of the SDSS using an adaptive matched filter  \citep{banerjee2018optical}. Figure \ref{fig:cluster-prop} shows the parameter distribution of galaxy clusters in this catalogue. It contains 46,479 galaxy clusters in the redshift range $0.045 \leq z \leq 0.641$ with a richness threshold $\lambda_{200} > 20$. Richness is a property related to the number of galaxies in a cluster and is defined as the total luminosity within the cluster radius  $R_{200}$, expressed in terms of luminosity ($L^*$). The cluster radius ($R_{200}$) used in this catalogue is the distance from the central galaxy within which the galaxy over density is 200 times the critical density of the universe ($\rho_{c}$). Based on this definition we can define a cluster mass $M_{200}$, where $M_{200}$ and $R_{200}$ satisfy the following relation:

\begin{equation}
    M_{200} = \frac{4\pi}{3} R^3_{200} \times 200 \rho_c
    \label{eq:mass-radius}
\end{equation}

\subsection{DESI galaxy cluster catalogue}

The DESI galaxy cluster catalogue (DGCC) is a catalogue generated from the   DESI legacy imaging surveys using a clustering algorithm, called Clustering by Fast Search and Find of Density Peaks \citep[CFSFDP,][]{zou2021galaxy}. It contains a total of 540,432 galaxy clusters in the redshift range $z \lesssim 1$. Figure \ref{fig:cluster-prop} also shows the parameter distribution of galaxy clusters from this catalogue. The characteristic cluster radius defined for this catalogue is $R_{500}$, defined analogously to the $R_{200}$, such that it satisfies the corresponding $R_{500}$-$M_{500}$ relation similar to the relation described in Eq. \ref{eq:mass-radius}.

\subsection{Catalogue uniformity}

The primary differences between the two catalogues described above are the richness threshold of $\ge20$ present in the SGCC and the respective definitions of cluster radius ($R_{200}$ for SGCC and $R_{500}$ for DGCC; Fig. \ref{fig:cluster-prop}). For consistency, we apply the same richness cut off of 20 to DGCC, and convert $R_{200}$ to $R_{500}$ for SGCC. Enforcing a richness threshold would also increase the purity of the clusters in DGCC, as it is noted from simulations that since no richness cut is applied while compiling it, the false detection rate can rise for low richness galaxy clusters \citep{zou2021galaxy}. We adopt $R_{500}$ as our definition of cluster radius to focus on SNe Ia in cluster cores, and to enable comparison with the results of \cite{ruppin2025ztf} and \cite{larison2024environmental}. The $R_{500}$ for SGCC is analytically computed using auxiliary parameters such as core radius and concentration provided in the catalogue, as detailed in the Appendix \ref{sec:r500_conversion}.

Each cluster catalogue was assembled using fundamentally different detection algorithms, therefore, despite enforcing uniform richness and cluster radius thresholds across both samples, we analyse the SN Ia population within each catalogue independently. For example, the SGCC assumes a  Navarro, Frenk \& White \citep[NFW,][]{nfw1, nfw2, nfw3} density profile and cluster properties are determined by parameter optimisation, whereas the DGCC is based on cluster identification through local density peaks and estimates $R_{500}$ and $M_{500}$ via empirical richness–mass scaling relation. These underlying methodological differences result in discrepancies in derived cluster parameters such $R_{500}$, as demonstrated in  Fig. \ref{fig:r500_comparison} for a few common clusters across both the catalogues. So, analysing SN Ia population within each catalogue separately prevents selection biases and catalogue-specific systematic uncertainties inherent in their construction.

\section{Methods} \label{sec:methods}

\begin{figure}[tbp]
    \centering
    \includegraphics[width=\linewidth]{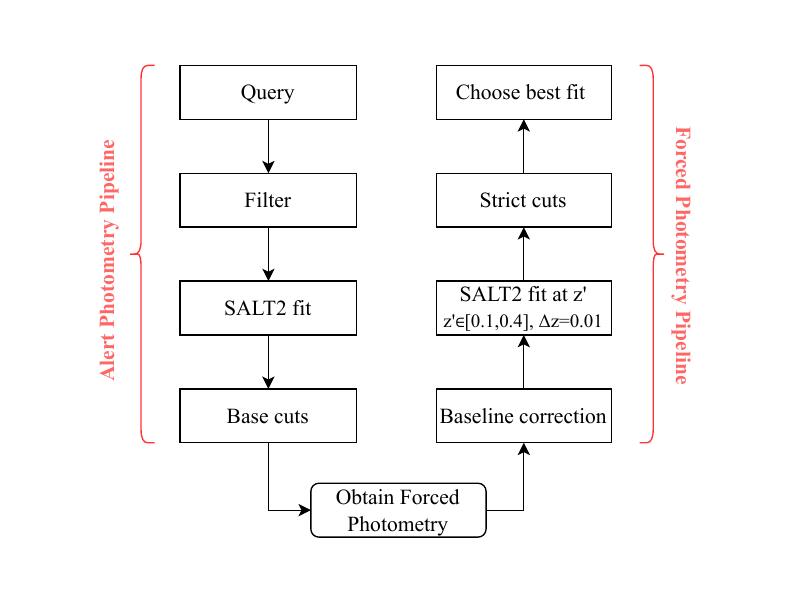}
    \caption{ AMPEL alert photometry and forced photometry pipeline described in Sect.  \ref{sec:ampel}. The initial query for spatially coincident  transients in galaxy clusters returns 89,025 unique ZTF objects from the archive, of which 11,087 remain after the base cuts. After processing the forced photometry light curves for these objects and the final selection cuts, we obtain the final sample of 4,969 likely SNe Ia used in this analysis. See Table \ref{tab:sne-cuts} for details on the base and strict cuts.  }
    \label{fig:ampel_work}
\end{figure}

This section details the process of compiling the galaxy cluster SN Ia sample, beginning with the extraction of transients from the ZTF archive, followed by our redshift estimation method and the SN Ia-cluster matching procedure.

\subsection{AMPEL pipeline: Compiling the SNe Ia sample} \label{sec:ampel}
To systematically curate a sample of transients in the ZTF archive spatially coincident with the galaxy cluster catalogues, we used the AMPEL framework\footnote{\url{https://ampelastro.github.io}} \citep{nordin2019transient}. AMPEL is an analysis and workflow framework for high throughput time-domain astronomy built with the core design goal of provenance tracking and reproducibility. AMPEL currently acts as a broker for ZTF and is also planned to operate as a broker for LSST \citep{nordin2025ampel}. Figure \ref{fig:ampel_work} illustrates the main stages of the alert photometry and forced photometry AMPEL workflow we have setup for this analysis.

\subsubsection{Querying ZTF archive}\label{sec:query}

\begin{figure}[tbp]
    \centering
    \includegraphics[width=0.9\linewidth]{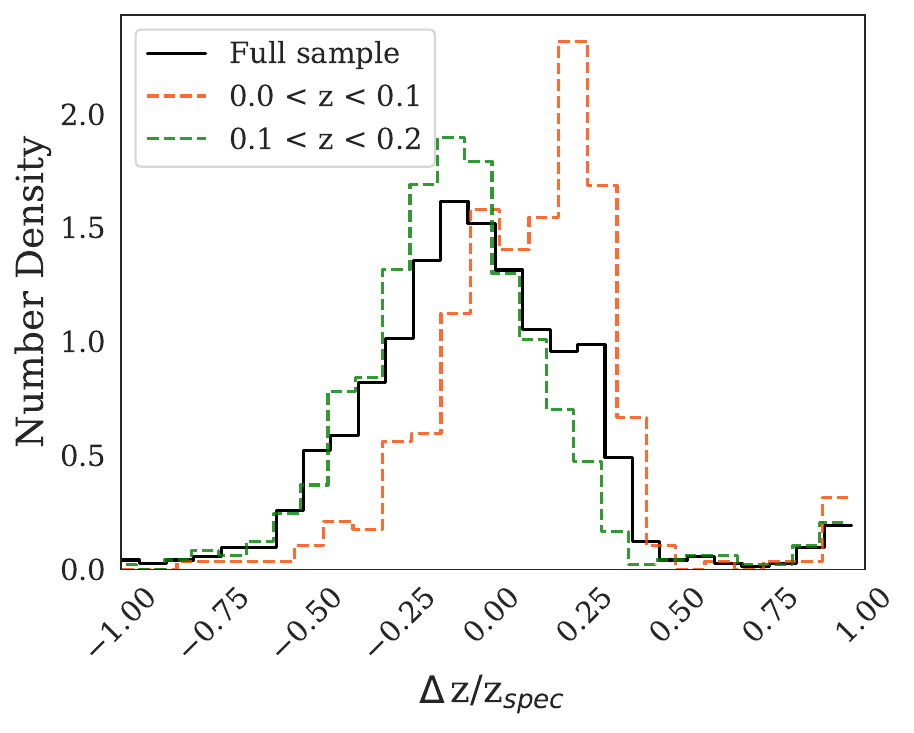}
    {\caption{(a)}}
    \ContinuedFloat
    \includegraphics[width=0.9\linewidth]{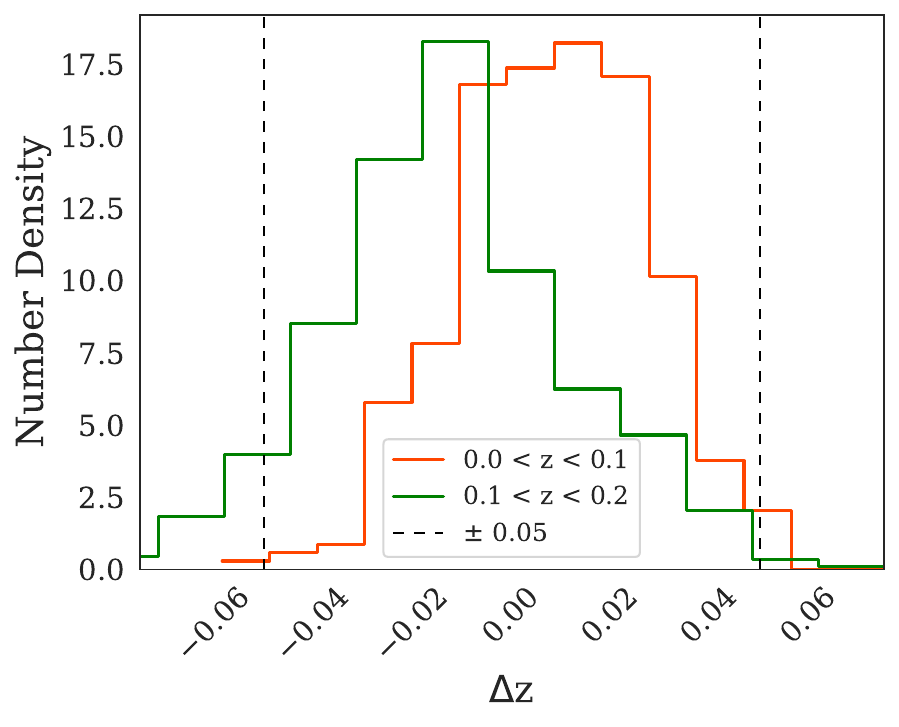}
    {\caption{(b)}}
    \ContinuedFloat
    \caption{ Relative error (a) and error (b) between the spectroscopic host redshift and estimated SN Ia photometric redshift using the multiple redshift bin fitting method described in Sect.  \ref{sec:strict} for $\sim22\%$ of our final 4,969 likely SNe Ia. $\Delta z$ is defined as $z_{spec} - z_{phot}$. }
    \label{fig:rel_error}
\end{figure}

Since ZTF is not a deep survey, we focus on galaxy clusters from the SGCC and DGCC in the redshift range $0.0\leq z\leq 0.2$. Combining these catalogues yields 4,852 galaxy clusters at $0.0\leq z\leq0.1$ and 25,255 galaxy clusters at $0.1\leq z\leq0.2$. For each galaxy cluster, we define the search radius as twice the characteristic cluster radius ($2R_{500}$) and query the archive for spatially coincident ZTF alerts. Querying the archive involves converting the search areas to their Hierarchical Equal Area and isoLatitude Pixelisation (HEALPix\footnote{\url{https://healpix.sourceforge.io}}) projection identifiers and ingesting the alert photometry. An initial set of selection criteria is applied during the query to reduce contamination from known stars (via cross match with Gaia and Pan-STARRS catalogue), and to distinguish genuine astrophysical events from false positives \citep[real bogus score,][]{2019MNRAS.489.3582D}. In addition, we use an alert time history of less than 60 days to reject long-lived transients such as tidal disruption events (TDE) and active galactic nuclei (AGN). These selections are described in the AMPEL unit \texttt{Decentfilter}\footnote{\url{https://github.com/AmpelAstro/Ampel-ZTF/blob/master/ampel/ztf/t0/DecentFilter.py}}. Furthermore, we require that each ZTF object (i.e. a persistent source associated with an alert) has at least five detections in the archive to ensure a well sampled light curve. This query returns a list of 89,025 ZTF objects. 

\subsubsection{Base cuts}
We applied an initial set of SN Ia cuts (base cuts) on the 89,025 ZTF objects based on the light curve  parameters obtained by fitting the data with a SALT 2.4 SN Ia template \citep{betoule2014improved} via the Python library \texttt{SNCosmo}  \citep{2016sncosmo}. The specific parameters used for this cut are listed in Table \ref{tab:sne-cuts}. This step brings down our sample to 11,087 preliminary SN Ia candidates. At this stage, the light curves  of these objects only consists of data points that exceed a fixed detection threshold from the ZTF automated processing pipeline and are distributed as public ZTF alerts \citep{masci2018zwicky}. To obtain more complete and better sampled light curves, we request forced photometry for these objects using the ZTF forced photometry service \citep{masci2023new}. Forced photometry measures the flux at the object's position on all of the difference images in the ZTF archive, irrespective of whether individual observation meets the detection threshold, thus producing a significantly more complete light curve.

\subsubsection{Redshift estimation}\label{sec:red_est}
The focus of this work is to maximise statistics by utilising all observed transients within the redshift range $z\leq0.2$. Estimating redshifts by cross-matching our sample with spectroscopic redshift catalogues, which are primarily limited to bright nearby objects, significantly reduces the sample size. Although photometric redshift catalogues offer broader coverage, Nordin et al. (in prep) demonstrated substantial offsets between spectroscopic and photometric redshifts for spectroscopically confirmed transients from the ZTF BTS \citep{2020ApJ...895...32F, 2020ApJ...904...35P}, particularly at redshifts $z\ge0.1$. Therefore, we do not infer redshifts for transients using catalogue cross-matching in this analysis. Instead, while processing the forced photometry light curves after the base cuts, we perform a series of SALT fits for each object across a range of redshift bins, from $z=0.01~\text{to}~0.4$ in steps of $\delta z=0.01$ and assign a redshift to the transient as the best-fit bin that passes the strict cuts described in the Sect. \ref{sec:strict}. We note that even if the exact redshift is not known, the light curve  fit quality should be optimal close to the true redshift. 

To validate our redshift estimation method, we selected a subset of our sample which has a spectroscopic host galaxy match from the DESI data release 1 catalogue\footnote{\url{https://data.desi.lbl.gov/doc/}} \citep{abdul2025data}. We subsequently perform a direct comparison between our estimated SN Ia redshift and the spectroscopic redshift of the corresponding host galaxies. The catalogue matching is carried out using the AMPEL catalogue matching service\footnote{\url{https://ampel.zeuthen.desy.de/api/catalogmatch/docs}} with a radius of 10 arcseconds. We find that only $\sim 22\%$ of our 4,969 likely SNe Ia have a host match from this catalogue. For this subset, Fig. \ref{fig:rel_error} shows the error and relative error in redshift between the estimated photometric redshift of the SN Ia and their spectroscopic host redshift. On further inspection, we find that the majority of objects with relative error $\Delta z/z_{spec} > 0.5$ have incorrect host associations; that is, either they are matched to a brighter galaxy near its host or the host galaxy spectrum does not exist. For objects with relative error $\Delta z/z_{spec}< -0.5$, the redshift estimate is incorrect and we overestimate the redshift of an actually closer object. Figure \ref{fig:rel_error}(b) demonstrates the consistency of our estimated photometric redshift with their spectroscopic host redshift, where, 
\begin{itemize}
    \item for $0.0 \leq z\leq 0.1: \sim 90\%$ of SNe Ia show an absolute error $\Delta z \leq 0.05$.
    \item for $0.1< z\leq 0.2: \sim 89\%$ of SNe Ia show an absolute error $\Delta z \leq 0.05$.
\end{itemize}

\subsubsection{Strict cuts}\label{sec:strict}
 We treat each light curve  fit within our redshift estimation framework (Sect. \ref{sec:red_est}) as a separate object and apply a second, more restrictive set of SN Ia cuts (strict cuts) listed in Table \ref{tab:sne-cuts}.
 In cases where fits at multiple redshift bins for a single ZTF object passes these strict cuts, the final selection involved retaining only the fit corresponding to the minimum reduced chi-squared ($\chi_{r}^2$). This gives us the final sample of 4,969 likely SN Ia candidates within the projected distance of $2R_{500}$.

\begin{table}[tbp]
    \centering
    \caption{Cuts applied at the alert photometry and forced photometry stages in our analysis to select likely SNe Ia. }
    \begin{tabular}{|c|c|c|}
    \hline
        Cut parameter  &Base cut  & Strict cut   \\
        \hline
         M$_{B}^{peak}$ & > -23  & -    \\
         M$_{B}^{std}$ &-& $\in$ [-20, -18.5]    \\
        \x& $\in$ [-5, 5]  & $\in$ [-4, 4]    \\
        c & < 4     & $\in$ [-0.2, 0.8]\\
        $\sigma_{x_1}$& < 5  & < 1.5   \\
        $\sigma_{c}$& < 0.5  & < 0.1   \\
        $\chi^2_{r}$& < 3 & < 2  \\
        
         \hline
    \end{tabular}
    \label{tab:sne-cuts}
    \tablefoot{M$_{B}^{peak}$ is the peak B-band absolute magnitude, M$_{B}^{std}$ is the standardised absolute magnitude. $x_1, c,\sigma_{x_1}, \sigma_{c}$ are the fit parameters from SN Ia template fitting.}
\end{table}

\subsection{Cluster-SN Ia matching}\label{sec:cl-sn}

Previous studies utilising spectroscopically classified SNe Ia defines a galaxy cluster-SN Ia association by calculating a membership probability based on redshift and redshift uncertainty of the  cluster and SN Ia as described in \cite{xavier2013properties}. As our sample is not spectroscopically identified and since only  $\sim 22\%$  have spectroscopic host redshifts we do not follow a similar approach and rather rely on the statistics offered by the large number of identified cluster SNe Ia. To identify a SN Ia located within galaxy clusters, we follow the steps below:

\begin{itemize}
\item Spatial coincidence: Match SN Ia candidates that are spatially coincident within twice the radius of the galaxy cluster ($2R_{500}$).
\item Redshift compatibility: We retain only those candidates whose estimated redshift match the cluster redshift within a tolerance window of $\Delta z_{\text{sep}}\leq0.05$, where $\Delta z_{\text{sep}} \equiv |~z_{\text{cluster}} - z_{\text{SN~Ia}}~|$.
\end{itemize}
 
 The SN Ia candidates are matched with the galaxy cluster catalogues using the \texttt{search\_around\_sky}\footnote{\url{https://docs.astropy.org/en/stable/api/astropy.coordinates.search_around_sky.html}}  function available in the \texttt{astropy} \citep{price2018astropy} library. To go from spatially coincident SNe Ia to a true SN Ia-cluster member match, we need to differentiate between foreground, background, and cluster SNe Ia. For this we define a redshift compatibility criterion of $\Delta z_{\text{sep}} \leq 0.05$, i.e., SN Ia candidates matched to a galaxy cluster are considered physically associated if they satisfy $|~z_{\text{cluster}} - z_{\text{SN~Ia}}~|\leq0.05$. This selection is informed by our analysis in Sect. \ref{sec:red_est} and illustrated in Fig. \ref{fig:rel_error}b, where utilising a subset of our sample with available spectroscopic host redshifts, we verified that our estimated photometric redshifts for $\sim 90\%$ of SNe Ia candidates fall within the $\Delta z \pm 0.05$ range. This relatively generous selection based on transverse cluster distance and redshift compatibility allows us to maximise the number of cluster SNe Ia. The noise introduced by this process, whether through incorrect association of a foreground or background SN Ia as a cluster member, or through misidentification of a transient as a SN Ia, is expected to be independent of the transverse cluster distance used in our analysis. We further try to minimise the noise by statistically removing chance foreground and background SN Ia, as detailed in Sect. \ref{sec:inter}. 

\subsection{Sample division}\label{sec:division}
The identified cluster SNe Ia are divided into two categories, core cluster SNe Ia which lie within the 50$\%$ radius of the galaxy cluster ($0.5R_{500}$), and outer-core cluster SNe Ia that lie outside this radius ($0.5R_{500}$ to $2R_{500}$). The analysis choice of using the radius as 50$\%$ is made such that it is large enough to have sufficient samples per bin for the analysis while small enough to minimise noise. We note that the general trends we observe do not change if a different cut-off radius is used.  We have identified 1,361 cluster SNe Ia, of which  107 (137) are core cluster SNe Ia and 487 (755) are outer-core cluster SNe Ia in the redshift range $z = 0.0-0.1 \, (0.1-0.2)$, respectively. The remaining SNe Ia in our sample ($\sim$3000), which passed all selection criteria but lack a cluster association following the matching procedure described in the previous section are treated as a reference population originating from field galaxies. We utilise this sample to compare against the cluster SN Ia distributions.

\subsection{Correcting for interloper contamination} \label{sec:inter}
\begin{figure*}[h!]
    \centering
    \includegraphics[width=\linewidth]{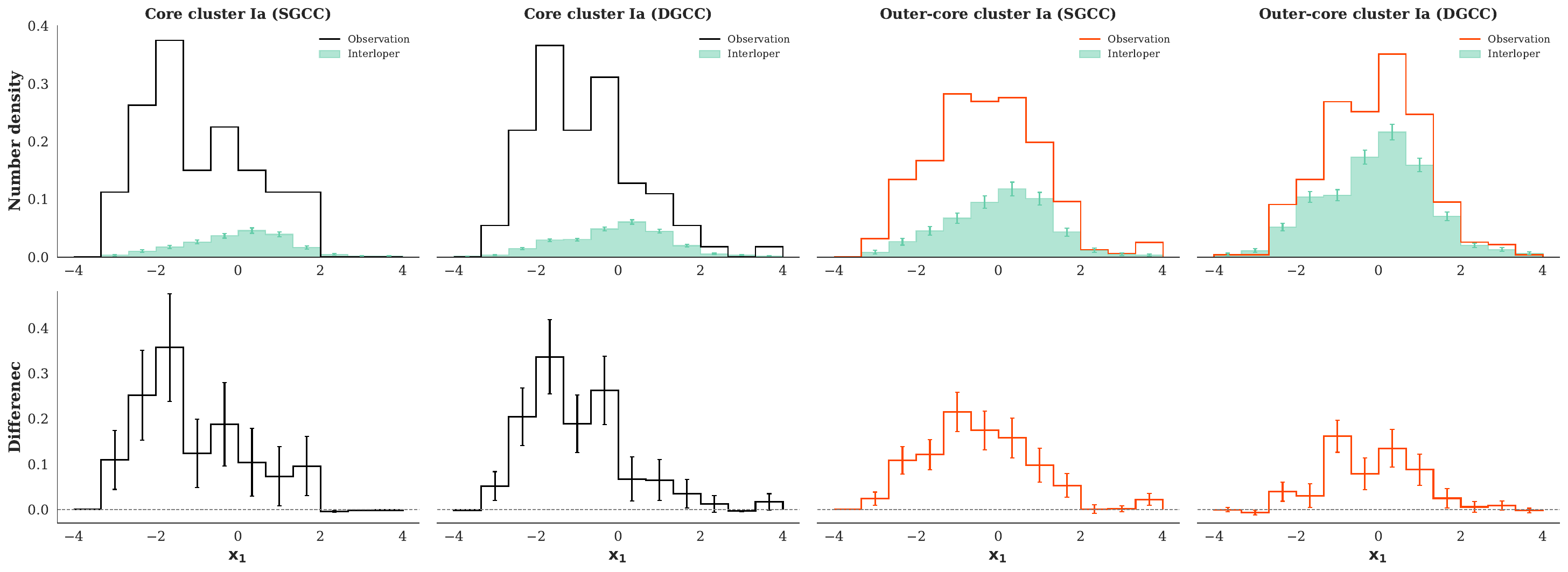}
    \caption{The top panel shows the observed stretch distribution of core cluster SNe Ia, outer-core cluster SNe Ia, and the computed interloper contribution (green) in the redshift bin $0.0\leq z<0.1$. The bottom panel displays the interloper corrected stretch distribution obtained by bin-by-bin subtraction of the interloper contribution from the observed sample counts.}
    \label{fig:subtraction0}
\end{figure*}

\begin{figure*}[h!]
    \centering
    \includegraphics[width=\linewidth]{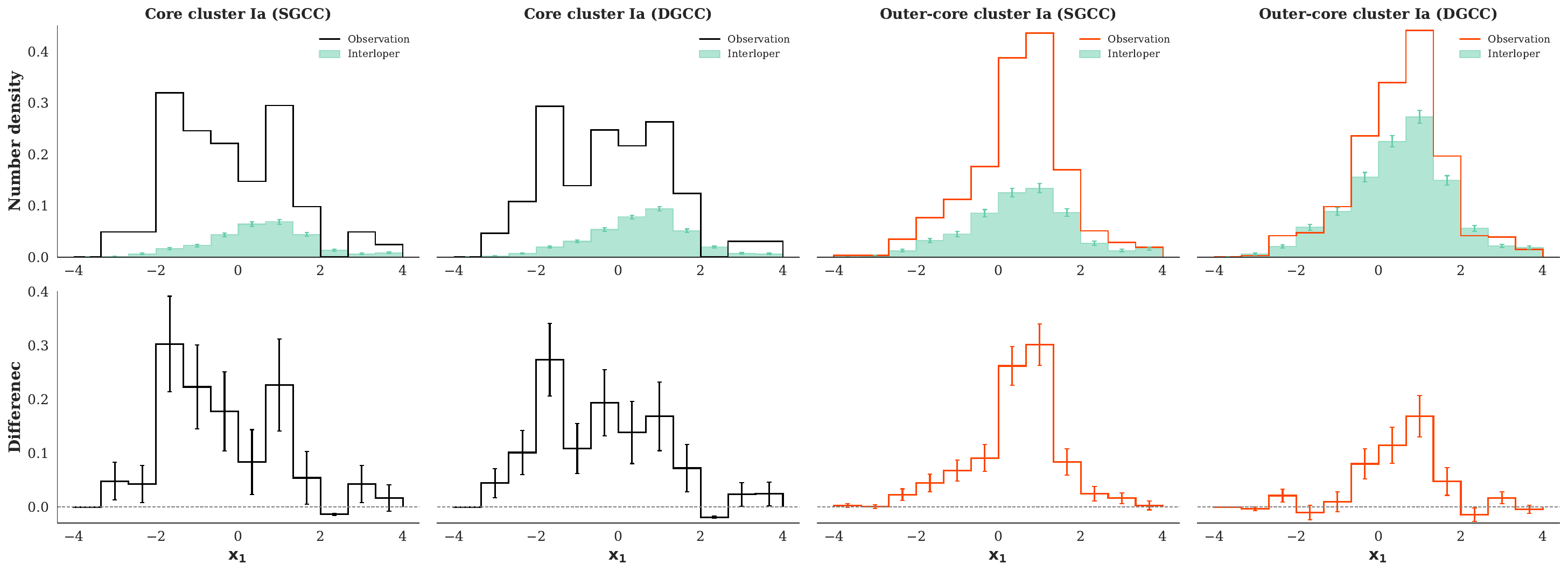}
    \caption{ Similar to Fig. \ref{fig:subtraction0} but for the redshift bin, $0.1\leq z<0.2$.    }
    \label{fig:subtraction1}
\end{figure*}

The cluster-SN~Ia matching (Sect.~\ref{sec:cl-sn}) allows a redshift offset of $\Delta z_{\text{sep}} = 0.05$ between a SN~Ia and its matched cluster. Since this window spans a large fraction of each redshift bin, some matched SNe~Ia are not physically associated with their cluster at all, but are foreground or background objects that happen to fall within the search aperture and, by chance, within the redshift window. We refer to these as interlopers, and correct for them using the local density of SNe~Ia far from the cluster centres, where any matches must be due to chance alignment.

For each cluster we define a reference shell beyond the core and outer-core boundaries, where SNe~Ia found within the same $\Delta z_{\text{sep}}$ window are by construction interlopers, since they share the same search geometry and selection function as the core/outer-core samples. We test different inner radii $r_{\text{ref}}$ for this shell, extending it from $r_{\text{ref}}=2R_{500}$ out to $r_{\text{max}} = 10R_{500}$, and compute the resulting SN~Ia surface density,
\begin{equation}
    \label{eq:stacked_dp}
    \Sigma(R_i) = \frac{N(R_i)}{n_{\text{clusters}} \times \Delta A_i} ,
\end{equation}
by binning all matched SN Ia-cluster pairs into fixed-width $R/R_{500}$ bins and dividing each bin's count by the number of clusters and its annulus area. Table~\ref{tab:sn_denisty} lists how the inferred interloper fraction changes with $r_{\text{ref}}$, and Fig. \ref{fig:sn_density} shows the resulting $\Sigma(R)$ profile for the two redshift bins (0.0 -- 0.1 and 0.1 -- 0.2) used in this analysis. In both bins, $\Sigma(R)$ declines steeply from $R/R_{500}=0$ and only flattens to a stable plateau at $R/R_{500} \gtrsim 4$. The decline up to the plateau shows that real cluster SN Ia excess extends well beyond our core and outer-core  definitions, consistent with prior work that defines cluster membership out to $5\,R_{500}$ \citep[e.g.,][]{ruppin2025ztf}, and the plateau is a signature that a genuine, interloper dominated regime has been reached. We therefore take $r_{\text{ref}} = 4R_{500}$ as the inner edge of the reference shell. Table \ref{tab:sn_denisty} shows for $r_{\text{ref}} \gtrsim 4R_{500}$ the interloper fraction in the core is stable at $f_{\text{core}} \approx 10$--13\% across both redshift bins and catalogues which is evidence that the core result is robust, with 87--90\% of the confirmed core sample unexplained by chance coincidence. At the outer-core boundary, the interloper fraction is higher but similarly stable, $f_{\text{outer-core}} \approx 30$--38\%, consistent with the expected increase in field contamination at larger projected radius.

The reference shell's surface density is
\begin{equation}
    \Sigma_{\text{interloper}} = N_{\text{interloper}} / (n_{\text{clusters}} \cdot A_{\text{interloper}}) ,
\end{equation}
which gives the expected number of chance interlopers in the core and outer-core regions, as
\begin{equation}
    N_{\text{core}}^{\text{exp}} = \Sigma_{\text{interloper}} \cdot A_{\text{core}}, \quad
    N_{\text{outer-core}}^{\text{exp}} = \Sigma_{\text{interloper}} \cdot A_{\text{outer-core}} .
\end{equation}
A true cluster SN Ia distribution can be obtained by subtracting the expected interloper contribution from the observed core and outer-core SN Ia populations. For instance, the stretch distribution measured for SNe Ia in the reference shell serves directly as the empirical interloper stretch template,  subtracting this component scaled by the determined interloper fraction from the observed core stretch distribution yields the true, interloper corrected cluster core SN Ia stretch profile.
\begin{equation}
    \text{Signal}_{\text{core}}(x_{1,i}) = O_{\text{core}}(x_{1,i}) - \text{Interloper}_{\text{core}}(x_{1,i}) ,
\end{equation}
where,
\begin{equation}
    \text{Interloper}_{\text{core}}(x_{1,i}) = h_{\text{interloper}}(x_{1,i}) \cdot \left( \frac{N_{\text{core}}^{\text{exp}}}{N_{\text{interloper}}} \right) ,
\end{equation}
and $h_{\text{interloper}}(x_{1,i})$ is the stretch distribution of $N_{interlopers}$. This can be analogously done for the colour distribution as well. Figures~\ref{fig:subtraction0} and~\ref{fig:subtraction1} show the resulting interloper corrected stretch distributions for the core and outer-core cluster SNe~Ia in each redshift bin.

\section{SN Ia properties in cluster cores} \label{sec:prop}

\begin{figure*}[!htbp]
   \centering
    \includegraphics[width=0.95\linewidth]{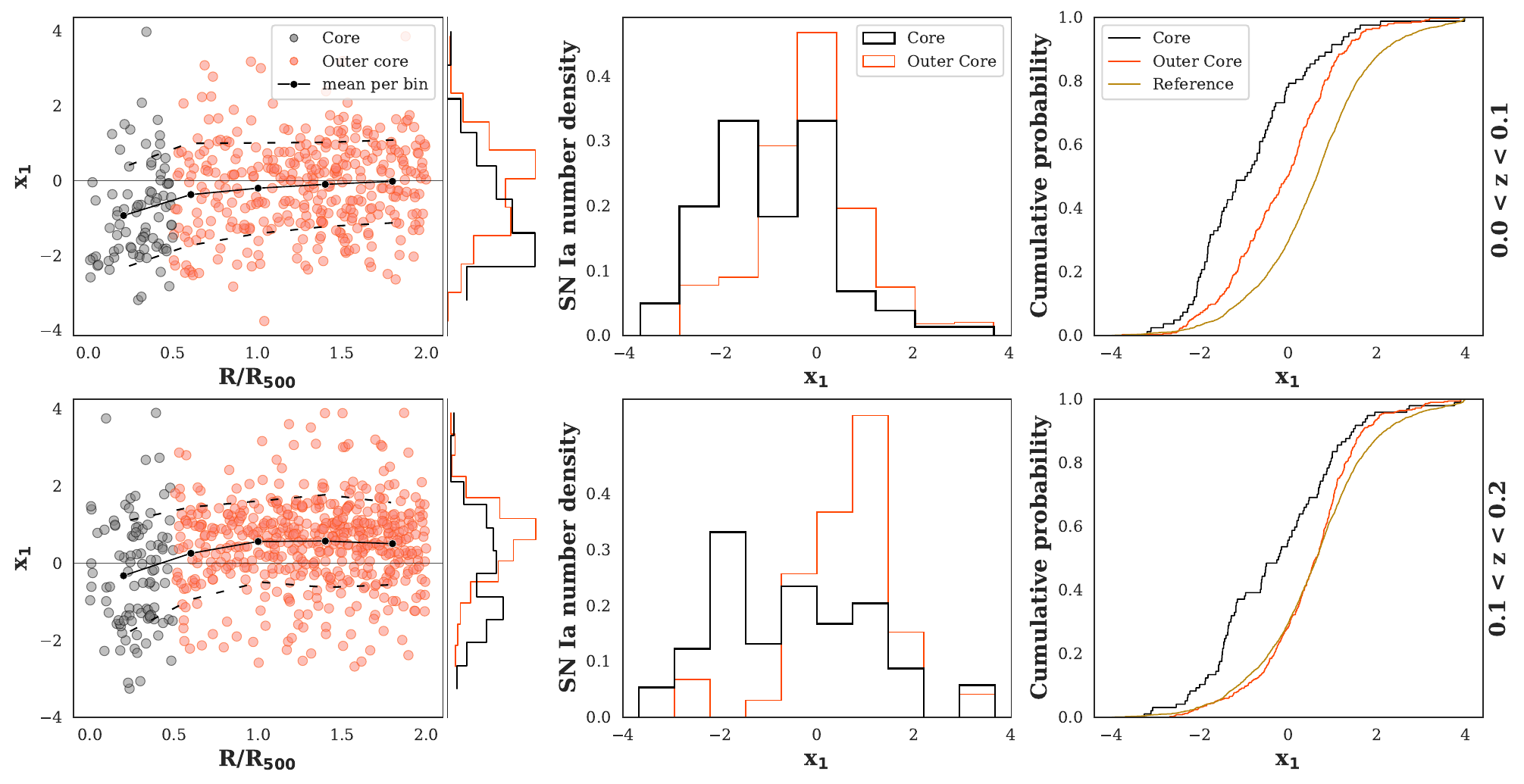}
      
      \caption{Stretch distribution of cluster SNe Ia derived from the DESI galaxy cluster catalogue. The top and bottom panels represent the redshift bins, $0.0\leq z\leq 0.1$ and $0.1< z\leq0.2$, respectively. The left plots shows the SN Ia stretch as a function of scaled projected distance from the galaxy cluster centre, with the running mean and standard deviation denoted by solid and dotted black lines. Adjoining side panels show the raw distributions for the core and outer-core cluster SNe Ia, while the middle plots display their corresponding interloper subtracted distributions. Right plots show the CDFs for the core and outer-core populations alongside the reference non-cluster population. A KS test comparing the core versus outer-core and core versus reference populations rules out the null hypothesis that they are drawn from the same parent distribution, with a significance of at least $4\sigma$ and $5\sigma$, respectively. A similar test between the outer-core and reference sample yields a p-values of $6.82 \times 10^{-18}$ for the low redshift bin and $5.3 \times 10^{-3}(\sim2.79\sigma)$ for the higher redshift bin.   }
      
     \label{fig:des_x1}   
   \end{figure*}

   \begin{figure*}[!htbp]
   \centering
   \includegraphics[width=0.95\linewidth]{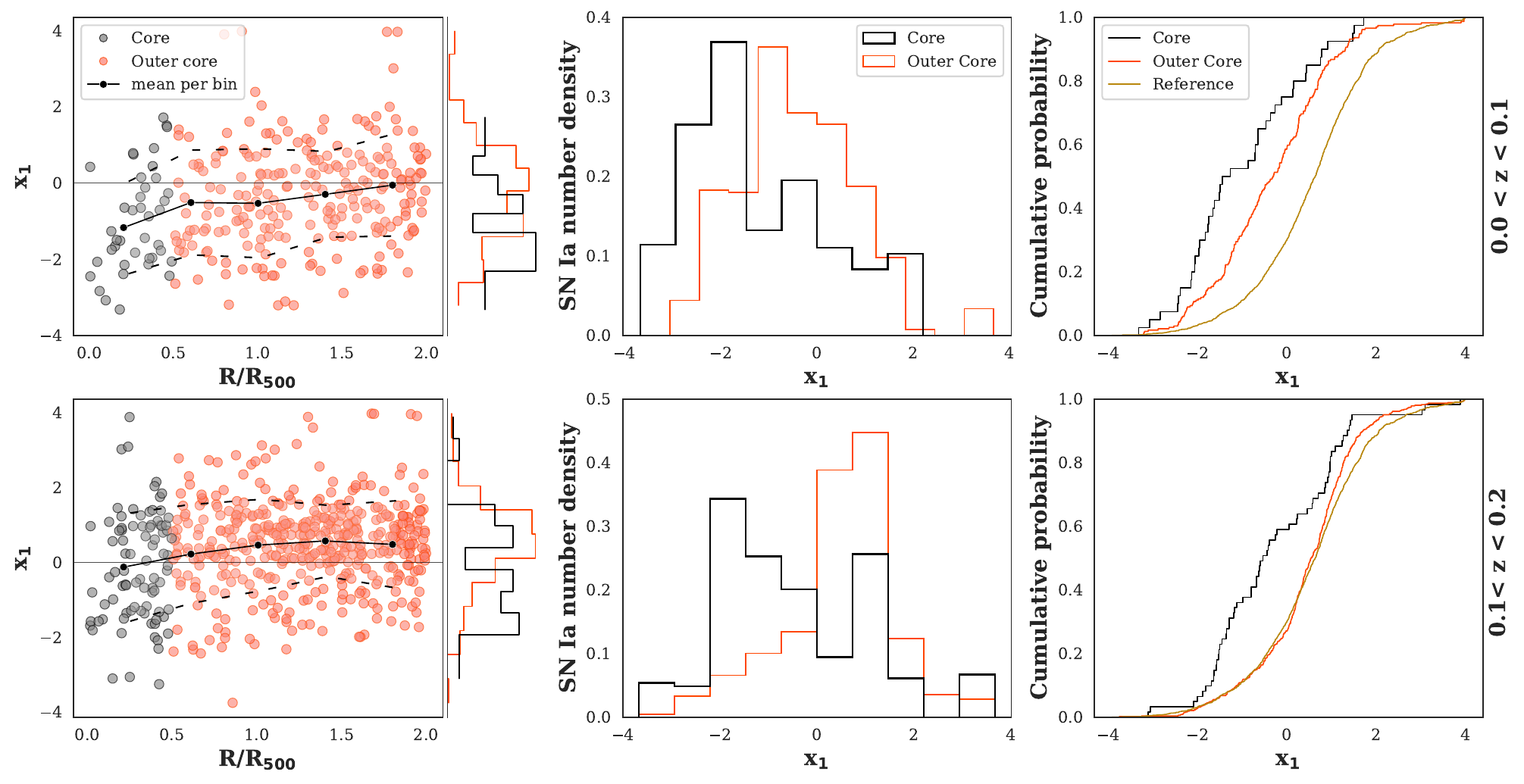}
      \caption{Stretch distribution of cluster SNe Ia  similar to Fig. \ref{fig:des_x1} but for the SNe Ia derived from the SDSS galaxy cluster catalogue. A KS test comparing the core versus outer-core and core versus reference population rules out the null hypothesis that they are drawn from the same parent distribution, with a significance of at least $3\sigma$ and $4\sigma$, respectively. Similarly the test between the outer-core and reference sample yields a p-values of $6.99 \times 10^{-17}$ for the low redshift bin and $1.53 \times 10^{-2}(\sim2.43\sigma)$ for the higher redshift bin.  
      }
     \label{fig:sdss_x1} 
   \end{figure*}

\subsection{Stretch distribution}

Figures~\ref{fig:des_x1} and \ref{fig:sdss_x1} show the stretch distributions of cluster SNe Ia scaled by cluster-centric distance, alongside the interloper corrected distributions of  core cluster  SNe Ia,  outer-core cluster SNe Ia, and their cumulative distribution functions (CDF). Across both the cluster catalogues and redshift bins ($0.0\leq z\leq 0.1$ and $0.1< z\leq0.2$), the stretch distribution of SNe Ia in galaxy cluster cores ($0.5R_{500}$) is systematically shifted toward more negative values ($x_1<0$), indicating a higher fraction of faster declining SN Ia population relative to the outer-core and reference SNe Ia population. A two-sided Kolmogorov-Smirnov (KS) test  on the light curve  stretch distributions of core and outer-core cluster SNe Ia rejects the null hypothesis that the two populations are identical with a significance of 4.41$\sigma$ (3.17$\sigma$) for the DGCC (SGCC) at $z \leq  0.1$. Similarly, for $z>0.1$, the null hypothesis is rejected at 5.54$\sigma$ (4.92$\sigma$) for DGCC (SGCC), respectively. Similar distinction is found when comparing the core and outer-core population against the reference distribution, yielding p-value consistent with zero across catalogues and redshift bins, with the only exception of outer-core versus reference comparison at $z>0.1$, where the statistical significance drops to $2.79\sigma$ for DGCC and $2.43\sigma$ for SGCC. 

Prior studies on cluster SNe Ia have found similar trends in their stretch distribution with SNe Ia in clusters declining faster than field SNe Ia \citep{xavier2013properties, toy2023rates, larison2024environmental, ruppin2025ztf}. These studies have focused on SNe Ia across large cluster radii, often extending to $5R_{500}$. In contrast, our study targets the more compact core region of $0.5R_{500}$. Combining this with the large sample of SNe Ia we have from the innermost core region, we can robustly test whether cluster SNe Ia are different. If progenitor host environments are indeed the primary driver of stretch diversity, the strongest signal for fast-declining SNe Ia should emerge from cluster cores. This is exactly what we observe with the stretch distribution of core cluster SNe Ia: a distinct population of fast-declining events. Furthermore, the fact that the core distribution  is statistically different from the outer-core and reference population of SNe Ia provides  strong empirical support for the two-population model, where SNe Ia have at least two principal environments. 

\begin{figure}[!htbp]
   \centering
   \includegraphics[width=0.95\linewidth]{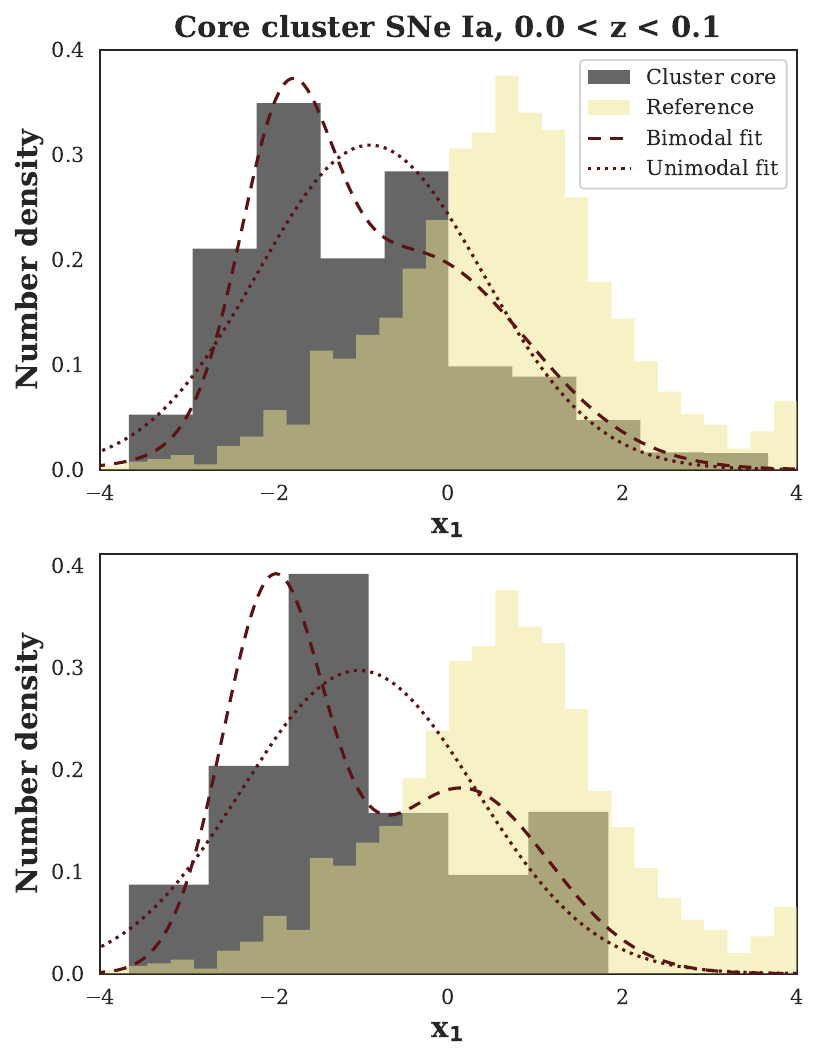}
    \caption{Bimodal Gaussian mixture model fit to the stretch distribution of core cluster SNe Ia at $0.0\le z \le 0.1$ for the two different galaxy cluster catalogues DGCC (top) and SGCC (bottom). The black histogram shows the stretch distribution of SNe Ia in cluster cores. The yellow histogram shows the stretch distribution from our non-cluster SNe Ia (field), dominantly composed of the slow-evolving (high \x) SNe Ia. The resulting fit parameters are listed in Table \ref{tab:fit_results}. The dotted lines show the unimodal fit to the data.}
     \label{fig:bimodal_fit_0}
\end{figure}

\begin{figure}[!htbp]
   \centering
   \includegraphics[width=0.95\linewidth]{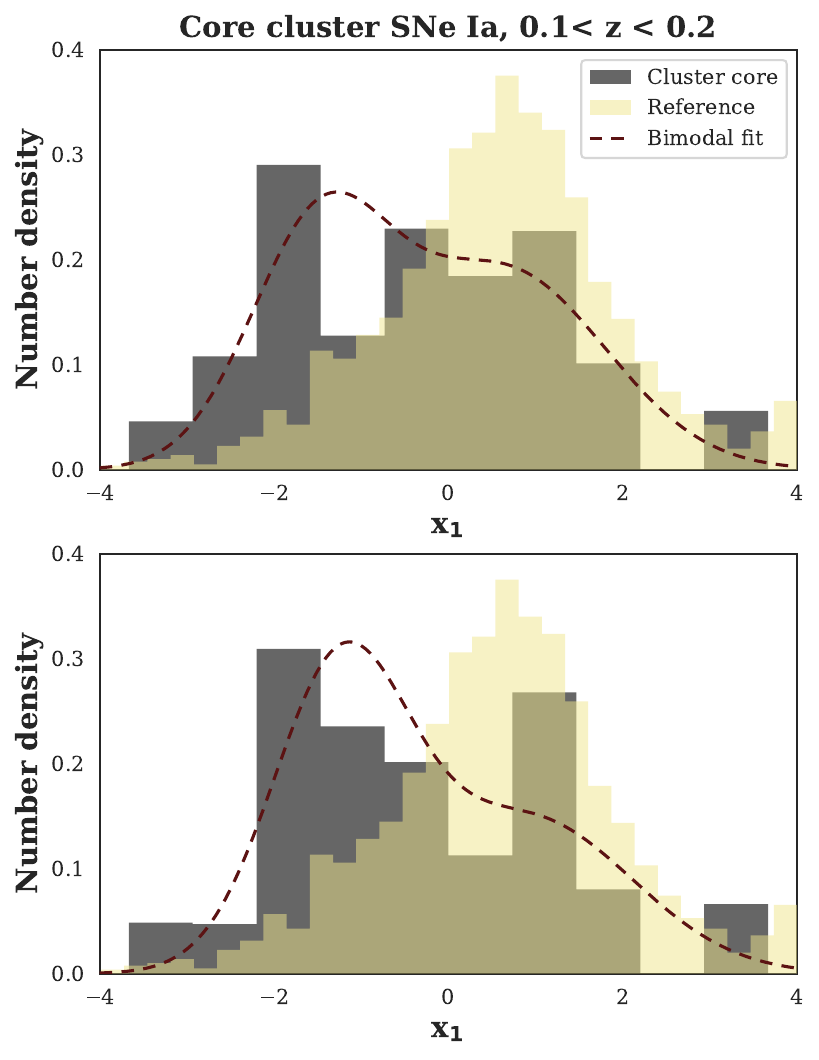}
    \caption{Bimodal Gaussian mixture model fit, similar to Fig. \ref{fig:bimodal_fit_0}, but for the redshift bin, $0.1< z \leq 0.2$. The best fit parameters are listed in Table \ref{tab:fit_results}. }
     \label{fig:bimodal_fit_1}
\end{figure}

If SNe Ia result from two populations, the stretch distribution should display two modes of varying amplitude, when studied as a function of cluster centric distance (going from core to outer-core distribution). At the same time, based on our methodology of photometrically identifying cluster SNe Ia, we would expect a contribution either from non-cluster SNe Ia present as contamination in our sample or normal SNe Ia occurring within the core. This population should exhibit a different distribution shape from the cluster core SNe Ia, with the relative amplitudes of these two populations varying with cluster-centric distance. To quantify this, we model the stretch distribution of core cluster SNe Ia (Fig. \ref{fig:bimodal_fit_0} and Fig. \ref{fig:bimodal_fit_1}) and outer-core cluster SNe Ia (Fig. \ref{fig:bimodal_fit_outer}) using a  bimodal Gaussian mixture model. The fit is executed using the \texttt{GaussianMixture}\footnote{\url{https://scikit-learn.org/stable/modules/generated/sklearn.mixture.GaussianMixture.html}} function from \texttt{scikit-learn}, where the free parameters, mean ($x_{1}^{low}$, $x_{1}^{high}$), dispersion ($\sigma_{low}$, $\sigma_{high}$), and amplitudes ($A_{low}, A_{high}$) are optimised via maximum likelihood estimation. The resulting best fit parameters for the core cluster distribution is listed in the Table \ref{tab:fit_results}. Ideally, the stretch distribution of the non-cluster SN Ia sample should constitute a bimodal distribution of SNe Ia originating from field galaxies, as observed with the ZTF DR2 volume-limited sample \citep{ginolin2025ztf}. However observational effects across our broad redshift ranges, such as Malmquist bias (the preferential detection of intrinsically bright objects in a flux limited survey), and the faster-fainter relation for SNe Ia (\citealt{1998A&A...331..815T}, where faster declining SN Ia are intrinsically fainter), makes our non-cluster sample dominated by high-stretch (slowly evolving) SNe Ia as seen in Fig. \ref{fig:bimodal_fit_0} and Fig. \ref{fig:bimodal_fit_1}.

\begin{figure*}[tbp]
   \centering
   \includegraphics[width=0.95\linewidth]{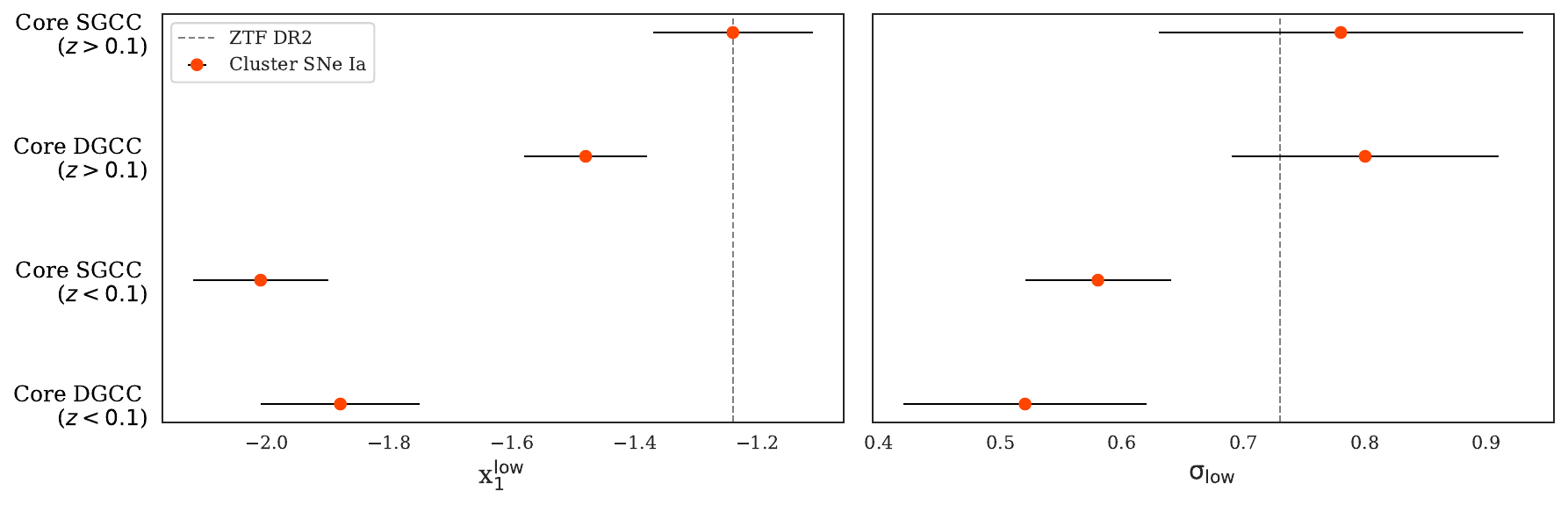}
    \caption{Mean ($x_1^{low}$) and standard deviation ($\sigma_{low}$) of the low-stretch mode in the bimodal Gaussian mixture fit to the  stretch distribution of SNe Ia in galaxy cluster cores. The value found for the low-stretch mode from the volume-limited field SNe Ia sample of ZTF DR2 is shown as vertical grey dotted lines. The best fit values of the model is listed in Table \ref{tab:fit_results}.  }
     \label{fig:x1_sigma}
   \end{figure*}

From the bimodal Gaussian mixture fit, we find that the mean of the low stretch mode for $0.0\leq z\leq0.1$ is $\bar{x}_{1}^{low} = -1.95\pm0.09$. This value is significantly lower than the low-stretch component observed in the ZTF DR2 field sample \citep[$x_{1}^{low}(ztf) = -1.24 \pm 0.18$;][]{ginolin2025ztf}, pointing to the unique predominance of fast-declining events within cluster cores (Fig. \ref{fig:x1_sigma}). Similarly, at $0.1<z\leq0.2$ the mean of the low stretch mode is $-1.36\pm0.08$, which is comparable to $x_{1}^{low}(ztf)$.
Since the core environment appears so overwhelmingly dominated by these fast-decliners, it raises the physical possibility that the core population is no longer inherently bimodal, but instead entirely explainable by a single, distinct mode of fast-declining events. To quantitatively test this hypothesis, we performed an alternative unimodal Gaussian fit to the core stretch distribution, and computed the Akaike Information Criterion (AIC) for both configurations ($\Delta AIC = AIC_{unimodal}-AIC_{bimodal}$) to determine whether the statistical complexity of a two-component model is justified by the data. We find that the bimodal model yields only a marginal improvement over the unimodal model for the DGCC ($\Delta AIC=2.3$), while for the SGCC ($\Delta AIC=1.53$), the unimodal model performs essentially as well as the bimodal configuration. Therefore, at $0.0\leq z\leq 0.1$, the data do not provide sufficient statistical leverage to confidently reject a unimodal distribution in favour of a bimodal one. This lack of clear bimodality hints that at the innermost cluster regions of $0.5R_{500}$, nearly the entire SN Ia population is comprised of a single, physically isolated population of fast-decliners.

A similar bimodal fit to the stretch distribution of the outer-core cluster SNe Ia is shown in Fig. \ref{fig:bimodal_fit_outer}. In contrast to the core region, the outer-core cluster sample does not exhibit a prominent low-stretch mode. Instead, its stretch distribution aligns closely with that of the non-cluster SN Ia sample. This lack of a distinct signal in the outer-core region is likely due to a decrease in sensitivity rather than a definitive physical absence owing to complex selection effects in our sample.

\subsection{Colour distribution}
A similar comparison of the SN Ia colour distribution shows no significant difference between the core cluster and outer-core cluster SNe Ia population. A two-sided KS test performed on these distributions fail to reject the null hypothesis, indicating that both samples are drawn from the same parent distribution with p-values of 0.89 (0.088) and 0.75 (0.49) for the lower and higher redshift bins, of DGCC (SGCC) respectively.  The colour distribution as a function of cluster centric distance, including the comparative histograms for the core and outer-core samples and their corresponding CDFs is shown in the Fig. \ref{fig:sdss_c}.  While SNe Ia exhibit some degree of intrinsic colour variation, the observed colour parameter is predominantly influenced by extrinsic factors, specifically extinction and reddening caused by dust within the host galaxy. Given that galaxy clusters are characterised by minimal dust content, we would not expect to observe significant colour variations between the core and outer-core cluster subpopulations. Our findings are consistent with this expectation.

\subsection{Selection effects}

The main selection effects in the analysis would be:
\begin{itemize}
    \item Contamination from non-SNe Ia transients that have passed all the SN Ia-like selection cuts.
    \item  Misidentification of a non-cluster SN Ia as a  cluster SN Ia.
\end{itemize}
 Our sample selection of photometrically identified SNe Ia relies on the observed homogeneity of SN Ia light curve, assuming that the vast majority of transients matching the SN Ia template and satisfying the strict quality cuts outlined in Table \ref{tab:sne-cuts} are indeed SNe Ia. An initial set of filters are already applied in the the ZTF archival query, to reduce contamination from known stars, TDEs and AGNs as mentioned in Sect.  \ref{sec:query}. To validate the purity of our selection, we cross-reference our SN Ia like cuts against the spectroscopically classified ZTF BTS catalogue. We perform SALT2 fits on $\sim 10,000$ BTS transients and find that fewer than $6\%$ of non-SN Ia transients pass our selection criteria. This demonstrates that the vast majority of our final sample of 4,969  should consist of genuine SNe Ia, confirming that the dataset is not subject to significant contamination from alternative transient classes.

Characterising the contamination from misidentification of a non-cluster SN Ia as a cluster member is complex, as it involves combination of multiple effects. For instance, when estimating the redshift of each SN Ia within our redshift-bin fitting methodology (Sect.  \ref{sec:red_est}), we operate under the assumption that the light curve  fit quality is usually best at or near the true redshift, and we use a cluster-SN Ia redshift compatibility criterion of $\Delta z_{\text{sep}}\leq0.05$. We made an initial attempt at reducing the contamination from non-cluster SN Ia  by statistically subtracting an interloper contribution from the observed sample (Sect. \ref{sec:inter}). In addition, to evaluate the reliability of our matching procedure, we tested how varying the criterion of  $\Delta z_{\text{sep}}\leq0.05$ impacts the final results. Although this threshold was motivated by the sub-sample of our dataset possessing spectroscopic host galaxy redshifts, we find that our conclusions remain highly stable under stricter constraints. Even when tightening this threshold to $\Delta z_{\text{sep}}\leq0.03$ and  $\Delta z_{\text{sep}}\leq0.02$ we consistently reject the null hypothesis that the core and outer-core SN Ia stretch distribution is drawn from the same parent distribution maintaining a statistical significance of at least the $3\sigma$ level, across both cluster catalogues and redshift bins. Additional effects would also come into play across our broad redshift range, as the sample completeness is not fully characterised and is likely subject to Malmquist bias, particularly at the higher redshift bin. While these factors inherently introduce potential uncertainties in both redshift determination and subsequent cluster-SN Ia association, given that contamination from non-Ia transients is already minimised (as demonstrated by our BTS cross-match), and an interloper distribution is subtracted, any residual misassociated SN Ia would act merely as an isotropic noise and is expected to be completely uncorrelated with the transverse cluster-centric distance. 

While we cannot identify a specific systematic artifact which could generate the observed fast-declining SNe Ia from cluster cores, we explicitly test the robustness of the observed dataset against contamination. We iteratively replace observed data points from the core cluster stretch distribution with randomly generated noise and map the exact threshold at which the KS test loses the statistical power to distinguish the core and outer-core populations.  Since the SN Ia stretch distribution natively displays a bimodal structure, the injected noise added to the core distribution was drawn from a bimodal Gaussian distribution matching the empirical features observed in the ZTF DR2 SN Ia sample \citep{ginolin2025ztf}.  As shown in Fig. \ref{fig:cont}, the two-sample KS test consistently rejects the null hypothesis at our strict significance threshold of $\alpha=0.01$ up to a high contamination fraction of $\sim45\%$. This indicates that even if nearly half of the core cluster SNe Ia consisted of background/foreground noise, the residual signal unique to the core population remains strong enough to be successfully resolved. Consequently, we can infer that the distinct nature of the core and outer-core cluster SNe Ia population is a genuine physical signal rather than a systematic selection artifact. 

\section{Discussions} \label{sec:disc}

\begin{table*}[!htbp]
\centering
\caption{ The resultant fit parameters from the bimodal Gaussian mixture model applied to the stretch distribution of core cluster SNe Ia from DGCC and SGCC (rounded to 2 decimal places).}
\label{tab:fit_results}
\begin{tabular}{lccccc}

Data & $x_{1}^{low}$ & $\sigma_{low}$ & $x_{1}^{high}$ & $\sigma_{high}$ & $A_{low}/A_{high}$ \\ \hline
\\
Core DGCC ($z \le 0.1$) & $-1.88 \pm 0.13$ & $0.52 \pm 0.10$ & $-0.39 \pm 0.17$ & $1.28 \pm 0.02$ & $0.52 \pm 0.23$ \\
Core SGCC ($z \le 0.1$) & $-2.01 \pm 0.11$ & $0.58 \pm 0.06$ & $0.15 \pm 0.06$ & $1.00 \pm 0.01$ & $1.20 \pm 0.02$ \\
Core DGCC ($z > 0.1$) & $-1.48 \pm 0.10$ & $0.80 \pm 0.11$ & $0.65 \pm 0.12$ & $1.16 \pm 0.18$ & $0.82 \pm 0.19$ \\
Core SGCC ($z > 0.1$) & $-1.24 \pm 0.13$ & $0.78 \pm 0.15$ & $0.93 \pm 0.15$ & $1.19 \pm 0.29$ & $1.28 \pm 0.24$ \\
\hline

\end{tabular}
\end{table*}

\subsection{Stretch distribution }

Our analysis of 242 cluster SNe Ia lying within $0.5R_{500}$ consistently recovers a population of fast-declining SNe Ia across the redshift range $0.0 \leq z \leq 0.2$ and two different optically selected galaxy cluster catalogues. The core sample exhibits a statistically significant difference in stretch distribution from both the outer-core and the reference non-cluster SNe Ia. The bimodal Gaussian mixture fit to the core stretch distribution finds $\bar{x}_{1}^{low} = -1.95\pm0.09$ across both catalogues for $z\le0.1$, much lower than the value typically observed for field SNe Ia \citep[e.g.,][]{nicolas2021redshift, ginolin2025ztf}. At $z\geq0.1$ the low-stretch mode is found at $\bar{x}_{1}^{low} = -1.24\pm0.18$, comparable to field values. For $z\le0.1$ we find no strong preference for a bimodal over a unimodal description of the core stretch distribution. These results shows that the SNe Ia in cluster cores are preferentially drawn from a low stretch population and host a significantly higher fraction of fast-declining events relative to the  outer cluster and field environments, with almost the entire core population at $z\le0.1$ explainable by a single low-stretch mode.

Cluster cores are dominated by old stellar populations with minimal star formation, so it is natural to expect the observed fast declining core cluster SNe Ia to arise predominantly from the oldest progenitor systems. This fits well with the two-population model of SNe Ia in which young star-forming and old quiescent galaxies share the high-stretch mode, while the low-stretch mode is exclusive to old environments, as suggested by \cite{rigault2020strong} and \cite{nicolas2021redshift}. However, we stress that this association is inferred from the environment and not measured directly through stellar-age estimates for the progenitor populations in this work. Studies have also established that the cluster cores are metal-rich relative to cluster outskirts ($R> 0.5R_{500}$) \citep{mantz2017metallicity, romero2023sami}. Disentangling the effects of progenitor age, metallicity and host mass remains a persistent challenge in SN Ia environmental studies and is beyond the scope of this work. Instead, our findings serve to empirically reaffirm the two-population model, demonstrating that there are at least two principal environmental channels producing SNe Ia and targetted observation at cluster cores can isolate the fast-declining population.

The following discussion is more model-dependent and should be regarded as a hypothesis motivated by our results.  While the two-population model is a reasonable explanation for what we observe, this study by itself cannot definitively determine whether the observed differences in stretch distribution between the core and outer-core cluster SNe Ia originate from two different progenitor channels or a single progenitor type, which under different explosion conditions can reproduce the entire range of SN Ia stretch diversity. However, cluster DTD studies suggest that the majority of SNe Ia in clusters at the redshifts considered in this work should arise from the DD system \citep{maoz2010supernova}. In that work the SD DTD gives a poor fit to their observed rate, primarily due to their inability to explain the delayed SNe Ia population, while purely DD models can only explain the rate only if the DD predictions are scaled up by factors 5-10. They find the DTD can be explained by a combination of prompt SD and delayed DD systems, with the delayed population being major contributor at $z<1.5$. If cluster cores suppress the prompt component, as expected for a quiescent environment, then the core cluster SNe Ia within our redshift range of $z\leq0.2$ would preferentially sample the delayed channel. Combining the cluster DTD finding with our observation of a distinct population of fast decliners in cluster cores makes the DD channel a viable candidate for the core population.

To discuss the plausible explosion mechanisms in such systems, we need estimates for $^{56}Ni$ and ejected mass, from detailed light curve  modelling and spectra for these SNe Ia, which cannot be achieved with our current photometrically identified sample. However, some works have tried to derive empirical relations between ejecta mass and stretch parameter based on bolometric light curve  modelling of a few normal SNe Ia \citep{scalzo2014type, scalzo2014ejected}. Applying equation~8 of \cite{scalzo2014ejected} to the mean stretch of the core population at $z<0.1$, with $x_1 = -1.95\pm0.09$, yields an ejecta mass of $0.92\pm0.05\,M_\odot$. This should be treated with caution as this is very much model dependent and it is applied here to the population mean and not to individual objects with measured ejecta masses. Taken at face value, such an ejecta mass would point to sub-Chandrasekhar mass explosions, suggesting that the progenitor of the core cluster SNe Ia is not only a delayed DD system, but specifically involves sub-Chandrasekhar mass explosions. A similar preference against Chandrasekhar mass explosion for fast-declining events is reported in recent analysis of the ZTF DR2 SN Ia sample \citep{2026arXiv260422928M}. Within the sub-Chandrasekhar scenarios, double detonation is the most widely discussed mechanism in DD systems \citep{woosley1994sub, fink2010double, 2010ApJ...714L..52S, kromer2010double, woosley2011sub}. It involves an initial detonation of a helium shell, acquired through merger or accretion from a degenerate companion, whose shock wave compresses the white dwarf and triggers a second detonation that unbinds the star. Although sub-Chandrasekhar explosions can also occur in SD systems where the helium shell is accreted from a He-rich non-degenerate star, it is unlikely that SD channel constitute the majority of the fast-declining events observed in cluster cores within this redshift range, as previously discussed. We regard the sub-Chandrasekhar and double-detonation interpretations as hypotheses to be rigorously tested through uniform spectroscopic follow-up of core cluster SNe Ia, combined with detailed empirical estimates of their ejecta masses.

 Our analysis has isolated a population of delayed, fast-declining SNe Ia within galaxy cluster cores. A follow up study focused on the standardisation of this specific sample can have interesting cosmological implication. The Hubble tension is one the most significant open questions in modern cosmology, and one way to address it is to explore the progenitor systems of SN Ia and test if the standardisation relations are population dependent. Recent evidences suggest linear standardisations in SN Ia can be more complex than what have been using. For instance, the stretch standardisation from the ZTF DR2 SNe Ia revealed a non-linear stretch magnitude relation \citep{ginolin2025ztf}. \cite{rubin2026banana} assuming two core populations of SN Ia for the Union3.1 sample, found evidence for distinct stretch magnitude relations for different populations. Similarly based on a two-population modelling of SNe Ia, \cite{wojtak2025stretch} demonstrated a reduction in the significance of the Hubble tension. Therefore, the unique core population isolated in this work represents a potentially useful sample for testing these theories. Furthermore owing to the relatively simpler star formation history in clusters compared to the field, future large scale population studies on cluster SNe Ia could provide a framework to test the redshift drift of the stretch parameter as previously identified in the field galaxies \citep{nicolas2021redshift}. Although our sample spans a significant redshift range and could in principle be used to study the redshift evolution of SN Ia properties,  the lack of a completeness model for the sample and the presence of Malmquist bias, makes such an investigation beyond the scope of this paper. We anticipate that the eventual photometric ZTF data release 4 (DR4), classified using deep learning techniques \citep[e.g.,][]{townsend2026photometricclassificationsupernovaedetected} will yield a significantly larger sample of photometrically identified SNe Ia and their associated host galaxies which would be instrumental in carrying out these high precision studies. 

\subsection{Colour distribution}
While our analysis primarily focuses on the stretch distribution, the colour distribution results are consistent with previous studies, where we observe no significant difference in colour distribution between the core cluster and outer-core cluster SNe Ia populations. We note that SN Ia colour is partially degenerate with redshift and our analysis may lack the sensitivity required to detect subtle environmental colour variations, should they exist. However, the observed uniformity in colour is consistent with the low dust environment in galaxy cluster cores \citep{mcgee2010dust}. The intra-cluster medium (ICM) filled with hot, dense plasma, is efficient in destroying dust grains through thermal sputtering. This process accounts for the low dust content observed in cluster cores. Hence, this would mean that SNe Ia in galaxy clusters especially in the core, should experience less extrinsic reddening, populating the red end of the colour distribution less than the field samples, which is what we observe from the colour distribution in Fig. \ref{fig:sdss_c}.

\section{Conclusions} \label{sec:conclude}
 We analysed the largest sample of photometrically identified galaxy cluster SNe Ia to date, with 242 SNe Ia found within a projected distance of $0.5R_{500}$ from the cluster centre and 1,197 SNe Ia found between $0.5R_{500}-2R_{500}$, from the ZTF archival data, across the redshift range $0.0\leq z \leq0.2$. Compared to prior works, we targeted the innermost core region of clusters where we expect to find the oldest stars, and consequently a population of SNe Ia driven by the characteristic properties of this environment. The main conclusions from our work are:
\begin{enumerate}
    \item We consistently recover a population of low-stretch (fast-declining) SNe Ia concentrated within  cluster cores ($R\leq 0.5R_{500}$) across two different galaxy cluster catalogues. This population is shown to be statistically different ($\sim4\sigma$) from the outer-core ($0.5R_{500}\leq R\leq 2R_{500}$) and reference (non-cluster) SNe Ia in our analysis. 
    \item The mean low-stretch value ($\bar{x}_{1}^{low}$) across both catalogues for $z\le0.1$, is found to be $-1.95\pm0.09$,  a value significantly lower than the low-stretch mode found from field SN Ia distribution, indicating that this population is strongly coupled to the quiescent environment of the cluster cores. The prominence of the low-stretch mode in cluster cores and its relative absence in the outer-core cluster regions support the two-population model where the low-stretch component is exclusively produced by older progenitor systems.
    \item Based on the DTD results in galaxy clusters and empirical relations between stretch and ejecta mass from external sources, we suggest that the cluster core SNe Ia originate from delayed DD systems involving sub-Chandrasekhar mass explosions, possibly via the double detonation scenario.
    \item Unlike stretch, SN Ia colour remains consistent across core and outer-core cluster environments. This uniformity is physically consistent with the efficient destruction of dust by the hot, dense plasma of ICM through thermal sputtering, leading to a sample with minimal extrinsic reddening. Though potential environmental colour variations may be masked by the inherent degeneracy between colour and our estimated redshift, this result remains consistent with prior works on cluster SNe Ia.
    \item Observations targeted at the innermost cluster cores ($0.5R_{500}$) of galaxy clusters can isolate the delayed channel of SNe Ia. This sample could be utilised to test SN Ia standardisation relations. Refining these relations for specific progenitor populations could be instrumental in addressing the Hubble tension and understanding the evolution of SN Ia properties across redshift.
\end{enumerate}

The results of this study highlight the power of leveraging wide-field surveys to carry out large scale population studies with photometrically identified SNe Ia. Future studies utilising the ZTF DR4 and Vera Rubin Observatory data will provide the statistical depth needed to further constrain the physics of these standardisable candles and their impact on precision cosmology.

\begin{acknowledgements}
This work was funded by the Deutsche Forschungsgemeinschaft (DFG, German
Research Foundation) – project number 460248186 (PUNCH4NFDI). Based on observations obtained with the Samuel Oschin Telescope 48-inch and the 60-inch Telescope at the Palomar Observatory as part of the Zwicky Transient Facility project. ZTF is supported by the National Science Foundation under Grants No. AST-1440341, AST-2034437, and currently Award \#2407588. ZTF receives additional funding from the ZTF partnership. Current members include Caltech, USA; Caltech/IPAC, USA; University of Maryland, USA; University of California, Berkeley, USA; Cornell University, USA; Drexel University, USA; University of North Carolina at Chapel Hill, USA; Institute of Science and Technology, Austria; National Central University, Taiwan, German Center for Astrophysics, Germany, and OKC, University of Stockholm, Sweden. Operations are conducted by Caltech's Optical Observatory (COO), Caltech/IPAC, and the University of Washington at Seattle, USA. The ZTF forced-photometry service was funded under the Heising-Simons Foundation grant \#12540303 (PI: Graham). Y.-L.K. was supported by the Lee Wonchul Fellowship, funded through the BK21 Fostering Outstanding Universities for Research (FOUR) Program (grant No. 4120200513819) and the National Research Foundation of Korea(NRF) grant funded by the Korea government(MSIT) (RS-2022-NR070872; RS-2022-NR070525; RS-2026-25473561). T.E.M.B. is funded by Horizon Europe ERC grant no. 101125877. U.B is funded by Horizon Europe ERC grant no. 101125877. MG is funded by the Horizon Europe ERC grant No 101002652 (BayeSN; PI K. Mandel).

\end{acknowledgements}

\bibliographystyle{bibtex/aa} 
\bibliography{ref}   
\begin{appendix}

\twocolumn
\section{Computing $R_{500}$ from $R_{200}$ for SGCC} \label{sec:r500_conversion}

The SGCC assumes a NFW density profile, which is given by, 
\begin{equation}
\rho(r) = \frac{\rho_s}{\left(r/r_s\right)\left(1+r/r_s\right)^2}
\end{equation}
and the enclosed mass of an NFW profile has a closed form:
\begin{equation}
M(r) = 4\pi \rho_s r_s^3 \, m(x), \qquad m(x) \equiv \ln(1+x) - \frac{x}{1+x}, \qquad x \equiv \frac{r}{r_s}
\end{equation}
The catalogue provides, for each cluster: $R_{200}$, the core radius $r_c$ (used as the NFW scale radius $r_s$), and the concentration $c_{200} \,\text{defined as}\, R_{200}/r_c$. Since $R_{200}$ and $r_s$ together fully specify the NFW profile, $R_{500}$ can be computed directly. By definition, $R_{200}$ satisfies, 
\begin{equation}
\bar\rho(R_{200}) = \frac{M(R_{200})}{\frac{4}{3}\pi R_{200}^3} = 200\,\rho_{\rm crit}
\end{equation}
Substituting the mass formula and letting $c \equiv R_{200}/r_s$:
\begin{equation}
\frac{3\rho_s\, m(c)}{c^3} = 200\,\rho_{\rm crit}
\quad\Longrightarrow\quad
\rho_s = \frac{200\,\rho_{\rm crit}\, c^3}{3\, m(c)}
\end{equation}
By the same logic, with $x_{500} \equiv R_{500}/r_s$:
\begin{equation}
\frac{3\rho_s\, m(x_{500})}{x_{500}^3} = 500\,\rho_{\rm crit}
\end{equation}
Dividing the $R_{500}$ equation by the $R_{200}$ equation cancels $\rho_s$ and $\rho_{\rm crit}$:
\begin{equation}
\frac{m(x_{500})}{x_{500}^3} = \frac{500}{200}\cdot\frac{m(c)}{c^3} = 2.5\,\frac{m(c)}{c^3}
\end{equation}
This is a single equation in one unknown, $x_{500}$, depending only on the concentration $c$.
Since $m(x) = \ln(1+x) - x/(1+x)$ is monotonic and smooth, $x_{500}$ can be found via a 1D root-finder:
\begin{equation}
\frac{m(x_{500})}{x_{500}^3} - 2.5\,\frac{m(c)}{c^3} = 0
\end{equation}
Finally, converting back to a physical radius using the tabulated core radius, the $R_{500}$ can be estimated using the relation:
\begin{equation}
R_{500} = x_{500} \times r_c
\end{equation}

\begin{figure}[h!]
        \centering
        \includegraphics[width=\linewidth]{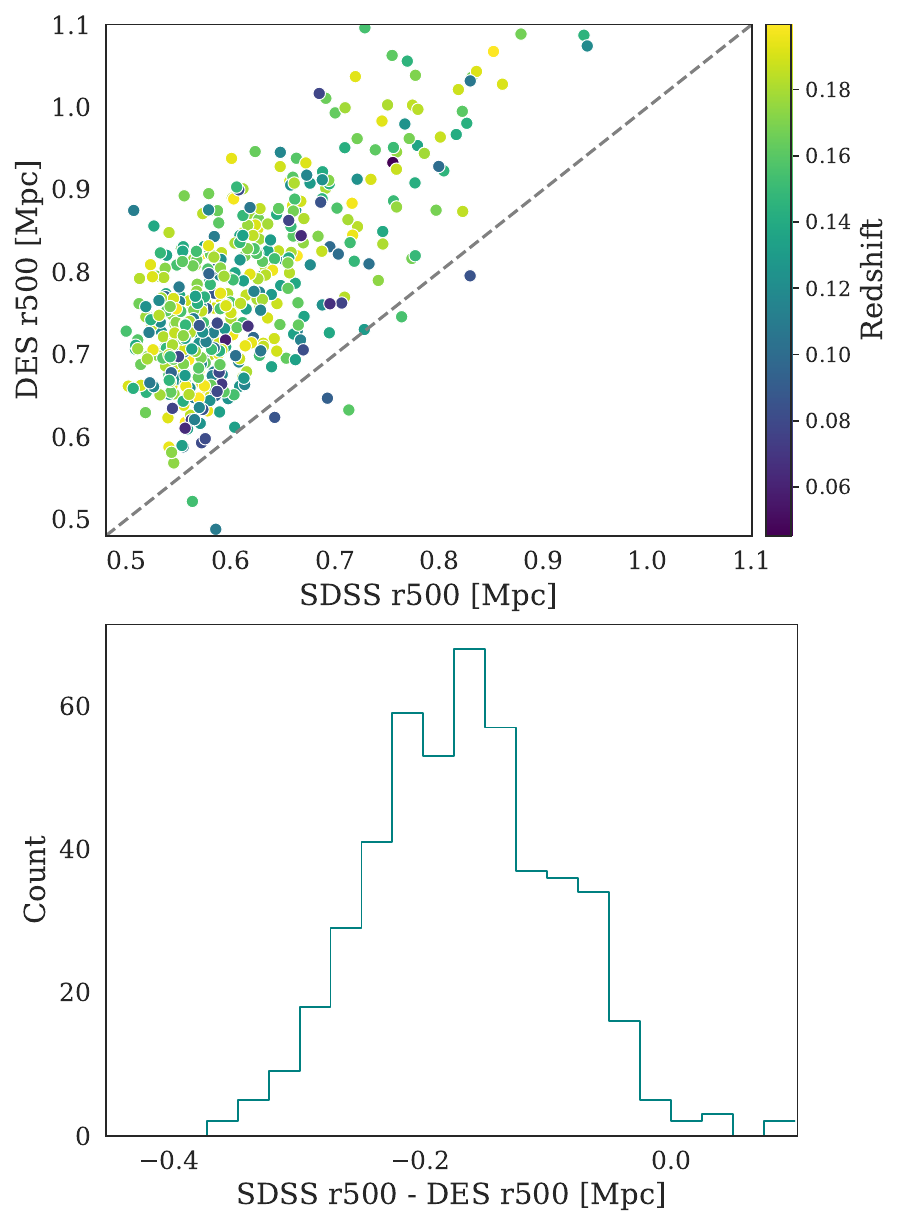}
        \caption{Comparison of $R_{500}$ values for common galaxy clusters in  DGCC (provided in the catalogue and derived using empirical richness–mass scaling relation) and SGCC (estimated from $R_{200}$ as described in this section).  }
        \label{fig:r500_comparison}
\end{figure}

Once we have standardised the characteristic cluster radius to $R_{500}$ across both catalogues, we can compare the estimated radii for common systems. Common clusters are identified across the two catalogues by requiring a  spatial separation within $30''$ and a redshift difference of $|\, \Delta\, z \,|\le0.02$. For the subset of clusters satisfying this criteria, Fig. \ref{fig:r500_comparison} shows the comparison between their $R_{500}$ values. The observed discrepancy in $R_{500}$ primarily arise from the distinct cluster detection algorithms employed by the two catalogues: SGCC adopts a parametric fit to compute $R_{500}$ under the assumption of an NFW radial density profile \citep{banerjee2018optical}, whereas DGCC derives $R_{500}$ from an empirical richness–mass scaling relation calibrated against X-ray and Sunyaev–Zel'dovich (SZ) observations \citep{zou2021galaxy}. Given this foundational algorithmic differences in construction of the catalogues, we analyse the SNe Ia populations in them independently. 

\FloatBarrier
\onecolumn

\section{Interloper subtraction and selection effects } \label{app:selection_eff}

\begin{table*}[!h]
\centering
\caption{Sensitivity of the inferred interloper fraction to the assumed reference shell radius, for each cluster catalogue and redshift bin.}

\label{tab:sn_denisty}
\resizebox{\textwidth}{!}{
\begin{tabular}{|cccccccccc|}
\hline
$r_{\text{ref}}$ & $r_{\text{max}}$ & $N_{\text{core}}$ & $N_{\text{outer-core}}$ & $N_{\text{interloper}}$ & $\Sigma_{\text{interloper}}$ & $N_{\text{core}}^{\text{exp}}$ & $N_{\text{outer-core}}^{\text{exp}}$ & $f_{\text{core}}$ & $f_{\text{outer-core}}$ \\
\hline
\hline

\multicolumn{10}{|c|}{\emph{DGCC $0.0 < z < 0.1$ }} \\
\hline
1.5 & 10.0 & 117 & 591 & 5374 & 0.004687 & 13.744246 & 206.163683 & 0.117472 & 0.348839 \\
2.0 & 10.0 & 117 & 591 & 5154 & 0.004577 & 13.421875 & 201.328125 & 0.114717 & 0.340657 \\
2.5 & 10.0 & 117 & 591 & 4957 & 0.004507 & 13.218667 & 198.280000 & 0.112980 & 0.335499 \\
3.0 & 10.0 & 117 & 591 & 4758 & 0.004457 & 13.071429 & 196.071429 & 0.111722 & 0.331762 \\
4.0 & 10.0 & 117 & 591 & 4337 & 0.004401 & 12.907738 & 193.616071 & 0.110323 & 0.327608 \\
5.0 & 10.0 & 117 & 591 & 3811 & 0.004332 & 12.703333 & 190.550000 & 0.108575 & 0.322420 \\
6.0 & 10.0 & 117 & 591 & 3182 & 0.004238 & 12.429687 & 186.445312 & 0.106237 & 0.315474 \\
7.0 & 10.0 & 117 & 591 & 2534 & 0.004236 & 12.421569 & 186.323529 & 0.106167 & 0.315268 \\
8.0 & 10.0 & 117 & 591 & 1787 & 0.004232 & 12.409722 & 186.145833 & 0.106066 & 0.314968 \\
\hline

\multicolumn{10}{|c|}{\emph{SGCC $0.0 < z < 0.1$ }} \\
\hline
1.5 & 10.0 & 40 & 234 & 2331 & 0.006789 & 5.961637 & 89.424552 & 0.149041 & 0.382156 \\
2.0 & 10.0 & 40 & 234 & 2243 & 0.006652 & 5.841146 & 87.617187 & 0.146029 & 0.374432 \\
2.5 & 10.0 & 40 & 234 & 2135 & 0.006484 & 5.693333 & 85.400000 & 0.142333 & 0.364957 \\
3.0 & 10.0 & 40 & 234 & 2019 & 0.006317 & 5.546703 & 83.200549 & 0.138668 & 0.355558 \\
4.0 & 10.0 & 40 & 234 & 1770 & 0.005999 & 5.267857 & 79.017857 & 0.131696 & 0.337683 \\
5.0 & 10.0 & 40 & 234 & 1561 & 0.005926 & 5.203333 & 78.050000 & 0.130083 & 0.333547 \\
6.0 & 10.0 & 40 & 234 & 1333 & 0.005930 & 5.207031 & 78.105469 & 0.130176 & 0.333784 \\
7.0 & 10.0 & 40 & 234 & 1084 & 0.006052 & 5.313725 & 79.705882 & 0.132843 & 0.340623 \\
8.0 & 10.0 & 40 & 234 & 757 & 0.005987 & 5.256944 & 78.854167 & 0.131424 & 0.336984 \\
\hline

\multicolumn{10}{|c|}{\emph{DGCC $0.1 < z < 0.2$ }} \\
\hline
1.5 & 10.0 & 162 & 867 & 9294 & 0.001599 & 23.769821 & 356.547315 & 0.146727 & 0.411243 \\
2.0 & 10.0 & 162 & 867 & 8936 & 0.001565 & 23.270833 & 349.062500 & 0.143647 & 0.402610 \\
2.5 & 10.0 & 162 & 867 & 8565 & 0.001536 & 22.840000 & 342.600000 & 0.140988 & 0.395156 \\
3.0 & 10.0 & 162 & 867 & 8137 & 0.001504 & 22.354396 & 335.315934 & 0.137990 & 0.386754 \\
4.0 & 10.0 & 162 & 867 & 7324 & 0.001466 & 21.797619 & 326.964286 & 0.134553 & 0.377121 \\
5.0 & 10.0 & 162 & 867 & 6432 & 0.001442 & 21.440000 & 321.600000 & 0.132346 & 0.370934 \\
6.0 & 10.0 & 162 & 867 & 5403 & 0.001420 & 21.105469 & 316.582031 & 0.130281 & 0.365147 \\
7.0 & 10.0 & 162 & 867 & 4253 & 0.001402 & 20.848039 & 312.720588 & 0.128692 & 0.360693 \\
8.0 & 10.0 & 162 & 867 & 2954 & 0.001380 & 20.513889 & 307.708333 & 0.126629 & 0.354912 \\
\hline

\multicolumn{10}{|c|}{\emph{SGCC $0.1 < z < 0.2$ }} \\
\hline
1.5 & 10.0 & 86 & 443 & 5107 & 0.002629 & 13.061381 & 195.920716 & 0.151877 & 0.442259 \\
2.0 & 10.0 & 86 & 443 & 4941 & 0.002590 & 12.867188 & 193.007812 & 0.149618 & 0.435684 \\
2.5 & 10.0 & 86 & 443 & 4752 & 0.002551 & 12.672000 & 190.080000 & 0.147349 & 0.429074 \\
3.0 & 10.0 & 86 & 443 & 4478 & 0.002476 & 12.302198 & 184.532967 & 0.143049 & 0.416553 \\
4.0 & 10.0 & 86 & 443 & 3894 & 0.002333 & 11.589286 & 173.839286 & 0.134759 & 0.392414 \\
5.0 & 10.0 & 86 & 443 & 3443 & 0.002310 & 11.476667 & 172.150000 & 0.133450 & 0.388600 \\
6.0 & 10.0 & 86 & 443 & 2929 & 0.002303 & 11.441406 & 171.621094 & 0.133040 & 0.387407 \\
7.0 & 10.0 & 86 & 443 & 2284 & 0.002254 & 11.196078 & 167.941176 & 0.130187 & 0.379100 \\
8.0 & 10.0 & 86 & 443 & 1552 & 0.002170 & 10.777778 & 161.666667 & 0.125323 & 0.364936 \\
\hline

\end{tabular}
}
\tablefoot{Each row corresponds to a different trial inner radius for the reference shell which extends from $r_{ref}$ to a fixed maximum radius, $r_{max} = 10\,R_{500}$. For each $r_{ref}$ the table list the number of matched SNe Ia within the cluster core ($N_{core}$, $R \leq 0.5\,R_{500}$, fixed across rows), cluster outer-core ($N_{\rm outer-core}$, $0.5\,R_{500} < R \leq 2\,R_{500}$, fixed across rows), and within the reference shell ($N_{\rm interloper}$, $r_{\rm outer} < R \leq r_{\rm max}$). The corresponding areal surface density of the reference shell, $\Sigma_{\rm interloper}$, is calculated using Eq. \ref{eq:stacked_dp}. $\Sigma_{\rm interloper}$, is then scaled by the core and outer-core areas to predict the number of interlopers expected in each region under that choice of reference shell boundary ($N_{\rm core}^{\rm exp}$, $N_{\rm outer-core}^{\rm exp}$). Dividing these by the observed counts gives the implied interloper fraction in each region ($f_{\rm core}$, $f_{\rm outer-core}$). A relative stable value of $f_{\rm core}, f_{\rm outer-core}$ across rows indicate the core and outer-core result is insensitive to the precise choice of $r_{\rm outer}$, whereas the $\Sigma_{\rm interloper}$ that decreases systematically with increasing $r_{\rm outer}$ shows that smaller boundary values still include genuine cluster members in the reference shell. The radius beyond which $\Sigma_{\rm interloper}$ stabilises is identified independently from the radial surface-density profile, and rows at or beyond that radius is most reliable as reference shell to estimate the interloper SNe Ia fraction.}
\end{table*}

\subsection{Interloper subtraction}

\begin{figure*}[h!]
        \centering
        \includegraphics[scale=0.55]{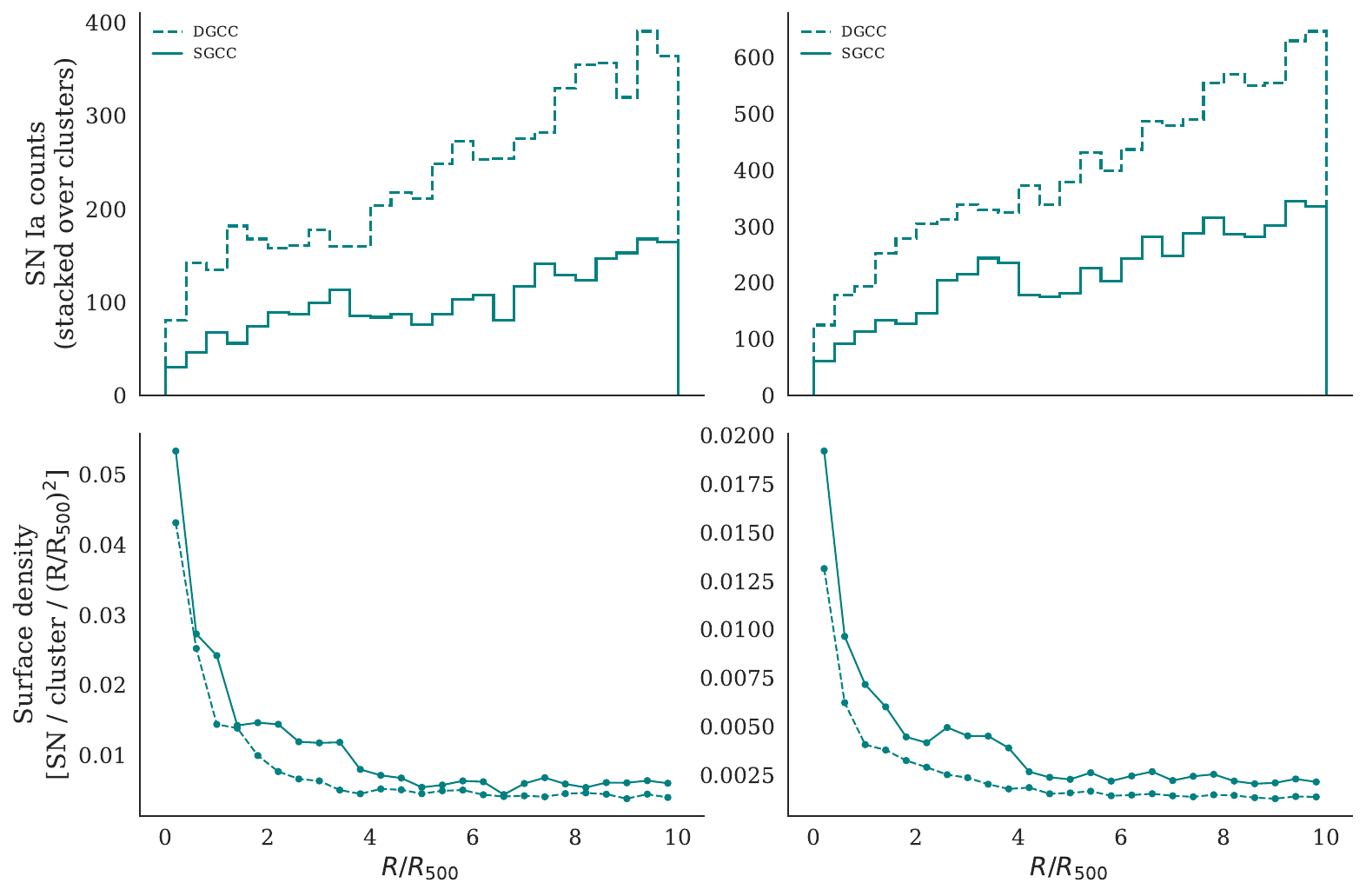}       
        \caption{Radial distribution of SNe~Ia around clusters for the DGCC (dashed) and SGCC (solid) catalogues, in the redshift bins $0.0 < z < 0.1$ (left) and $0.1 < z < 0.2$ (right). Top: SN~Ia counts stacked over all clusters as a function of cluster-centric distance, $R/R_{500}$. Bottom: the corresponding surface density (Eq.~\ref{eq:stacked_dp}), i.e.\ counts per cluster per unit area. The profile declines steeply from the cluster centre and flattens into a plateau at $R/R_{500} \gtrsim 4$, which we take as the onset of the interloper-dominated reference shell ($r_{\text{ref}} = 4\,R_{500}$).}
        \label{fig:sn_density}
\end{figure*}

\subsection{Noise injection and contamination sensitivity}
\begin{figure}[h!]
    \centering
    \includegraphics[scale=0.5]{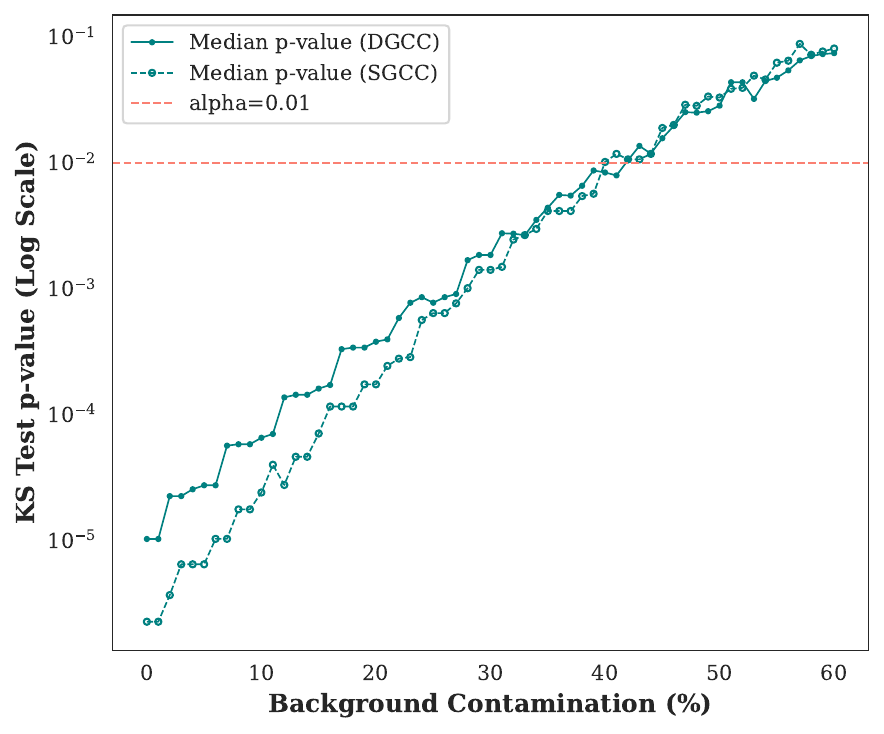}
    \caption{KS test significance threshold analysis under varying levels of bimodal noise injection within the stretch distribution of core cluster SNe Ia, for the redshift bin of $z\le0.1$.  }
    \label{fig:cont}
\end{figure}

\FloatBarrier
\onecolumn
\section{Supplementary figures for the analysis}

\subsection{Catalogue Projections}
\begin{figure*}[h!]
        \centering
        
        \includegraphics[scale=0.45]{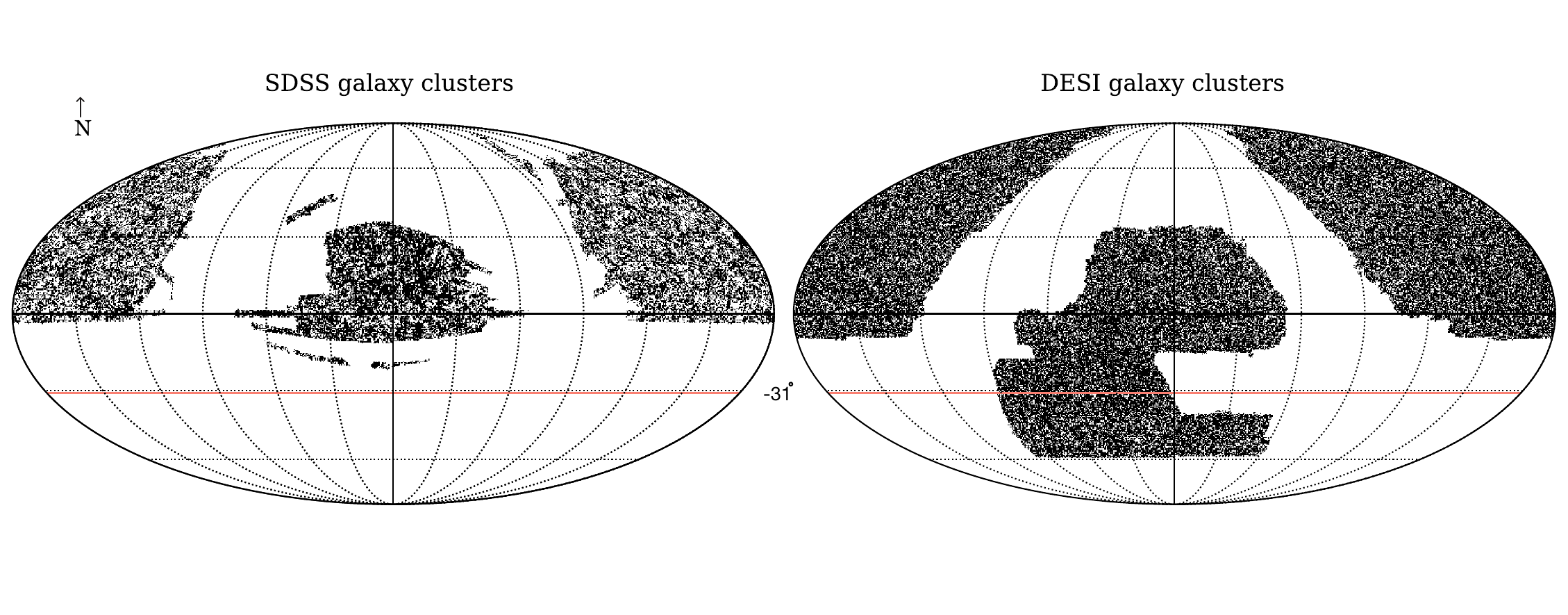}
        
        \caption{Mollweide projection of galaxy clusters for the SDSS (left) and DESI (right) galaxy cluster catalogues. The horizontal line shows the maximum declination towards south observable by ZTF. }
        \label{fig:projection}
\end{figure*}

\subsection{Cluster SNe Ia colour distribution and model fitting}
\begin{figure*}[h!]
   \centering
   \includegraphics[scale=0.5]{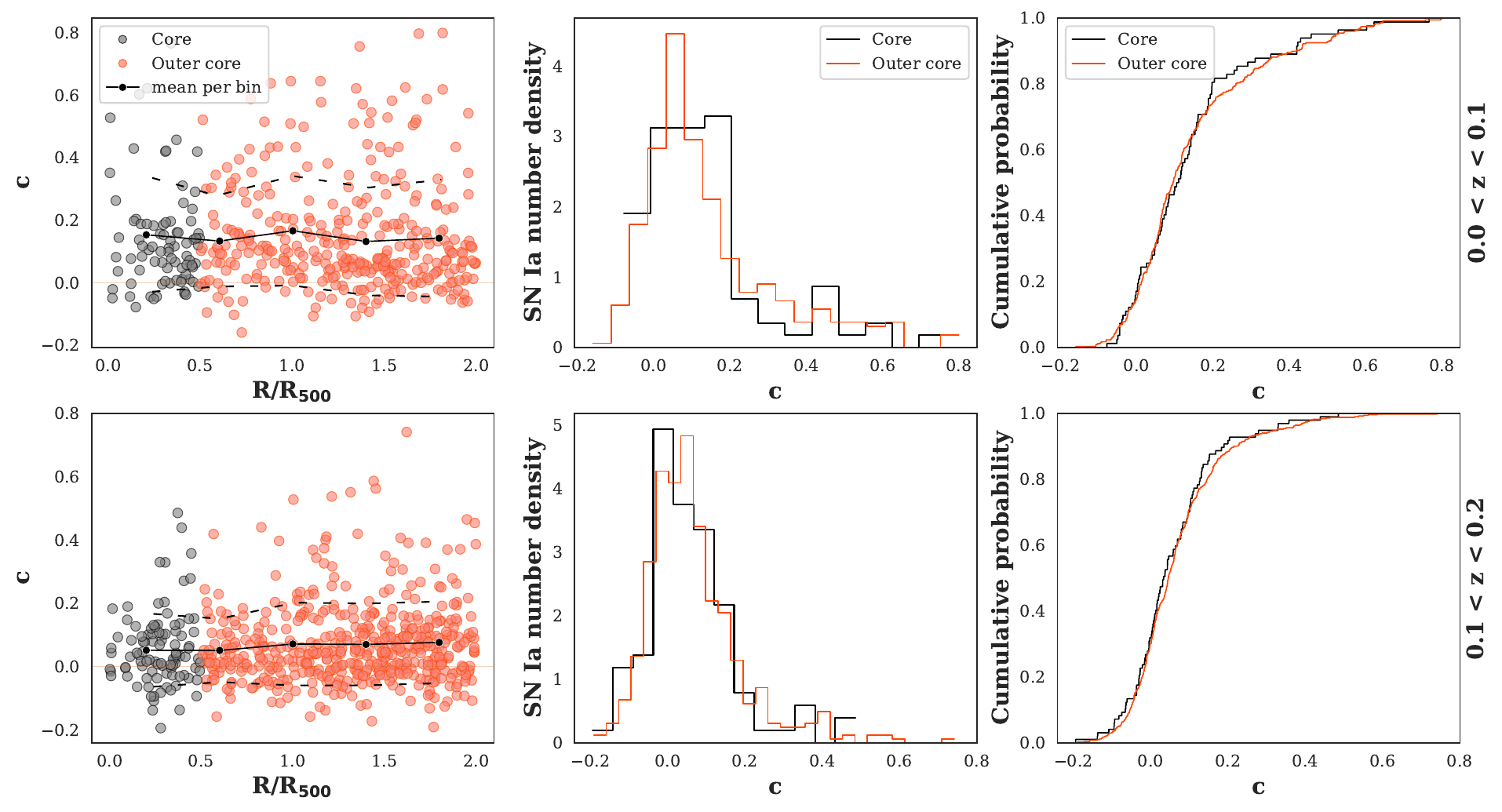}
      \caption{SN Ia colour distribution for cluster SNe Ia from the DESI galaxy cluster catalogue. The top and bottom panels are for the redshift bins, $0\leq z\leq 0.1$ and $0.1\leq z\leq0.2$, respectively. The left plot shows the SN Ia colour as a function of scaled projected distance to the galaxy cluster centre ($R/R_{500}$), with the running mean and standard deviation denoted by solid and dotted black lines. The middle plots show the colour distribution of core cluster and outer-core cluster SNe Ia, and the right-hand plots showing their CDFs along with the CDF of the reference non-cluster sample. A KS test on the core and outer-core SN Ia colour distributions support the null hypothesis that both samples are drawn from the same parent distribution ( p-values of 0.89 for the lower redshift bin and 0.75 for the higher redshift bin).
      }
         \label{fig:sdss_c}
   \end{figure*}

\begin{figure*}[h!]
   \centering
  \includegraphics[scale=0.5]{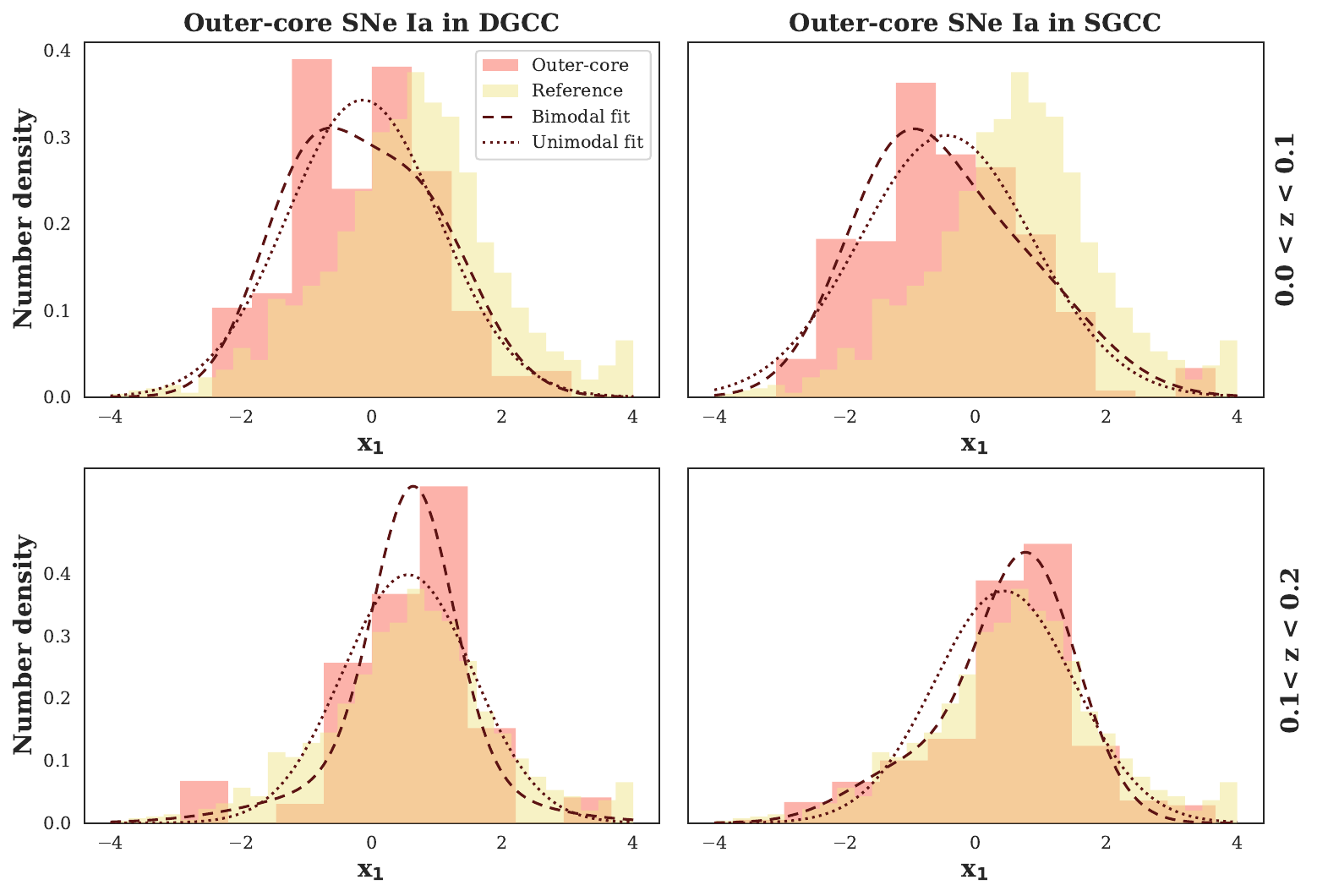}
    \caption{Bimodal Gaussian mixture fit to the stretch distribution of outer-core cluster SNe Ia from Fig. \ref{fig:des_x1} and Fig. \ref{fig:sdss_x1}.}
     \label{fig:bimodal_fit_outer}
   \end{figure*}

\end{appendix}
\end{document}